\documentclass[11pt]{article}
\title{Exact MLE for Generalized Linear Mixed Models}
\author{Tonglin Zhang\footnote{Department of Statistics, Purdue University, 150 North University Street,West Lafayette, IN 47907-2066, Email: tlzhang@purdue.edu} \\Purdue University}
\usepackage{epsfig}
\usepackage{mathrsfs}
\usepackage{amsfonts}
\usepackage{amsmath}
\usepackage{wrapfig,lipsum,booktabs}
\usepackage{bm}
\usepackage{algorithm}
\usepackage{algpseudocode}
\def\qed{\hfill$\diamondsuit$}

\newtheorem{thm}{Theorem}
\newtheorem{cor}{Corollary}
\newtheorem{lem}{Lemma}
\begin{document}
\maketitle
\def\eqalign#1{\null\,\vcenter{\openup\jot\ialign
              {\strut\hfil$\displaystyle{##}$&$\displaystyle{{}##}$
               \hfil\crcr#1\crcr}}\,}

\setcounter{page}{1}

\begin{abstract}

Exact MLE for generalized linear mixed models (GLMMs) is a long-standing problem unsolved until today. The proposed research solves the problem. In this problem, the main difficulty is caused by intractable integrals in the likelihood function when the response does not follow normal and the prior distribution for the random effects is specified by normal. Previous methods use Laplace approximations or Monte Carol simulations to compute the MLE approximately. These methods cannot provide the exact MLEs of the parameters and the hyperparameters. The exact MLE problem remains unsolved until the proposed work. The idea is to construct a sequence of mathematical functions in the optimization procedure. Optimization of the mathematical functions can be numerically computed. The result can lead to the exact MLEs of the parameters and hyperparameters. Because computing the likelihood is unnecessary, the proposed method avoids the main difficulty caused by the intractable integrals in the likelihood function. 

\end{abstract}



\section{Introduction}
\label{sec:introduction}

Generalized linear mixed models (GLMMs) are extended from generalized linear models (GLMs) when normally distributed random effects are involved in the linear components. They provide a Bayesian framework for modeling responses from exponential family distributions.  Except for a response following a multivariate normal distribution given the random effects, all the remaining cases of GLMMs contain intractable integrals in their likelihood functions, leading to difficulty in computing the exact MLEs of parameters and hyperparameters. Previous methods use the Laplace approximation or the MCMC techniques to overcome the difficulty. These methods can only provide the approximate MLEs, implying that the exact MLE problem remains unsolved. The proposed research solves the problem.

We assume a dispersion parameter is not involved in a GLMM because it is appropriate for binomial, Poisson, and multinomial data. The response vector ${\bm y}=(y_1,\dots,y_n)^\top\in\mathbb{R}^n$ of the GLMM can be expressed by an exponential family distribution as
\begin{equation}
\label{eq:exponential family distribution}
f({\bm y}|{\bm\gamma})=f_{{\bm\beta}}({\bm y}|{\bm\gamma})=\exp[{\bm y}^\top{\bm\eta}-{\bf 1}^\top b({\bm\eta})+{{\bf 1}^\top}c({\bm y})],
\end{equation}
where ${\bm\eta}=(\eta_1,\dots,\eta_n)^\top$ is an $n$-dimensional linear components vector, $b({\bm\eta})=(b(\eta_1),\dots,b(\eta_n))^\top$ is derived by a real transformation $b(\cdot)$ on ${\bm\eta}$,  and $c({\bm y})=(c(y_1),\dots,c(y_n))^\top$ is a normalized constant vector. A GLMM with the canonical link is
\begin{equation}
\label{eq:generalized linear mixed model}
{\bm\eta}={\bm\eta}_{{\bm\beta}{\bm\gamma}}={\bf X}{\bm\beta}+{\bf Z}{\bm\gamma}
\end{equation}
and 
\begin{equation}
\label{eq:distribution of random effects}
{\bm\gamma}\sim{\cal N}({\bf 0},{\bf D}),
\end{equation}
where ${\bf D}={\bf D}_{\bm\omega}$ is determined by a hyperparameter vector ${\bm\omega}=(\omega_1,\dots,\omega_{r})^\top\in\mathbb{R}^r$ for the variance components, ${\bf X}=({\bm x}_1^\top,\dots,{\bm x}_n^\top)^\top\in\mathbb{R}^{n\times p}$ with ${\bm x}_{i}=(x_{i0},\dots,x_{i(p-1)})^\top\in\mathbb{R}^p$ is a design matrix for fixed effects, ${\bm\beta}=(\beta_0,\dots,\beta_{p-1})^\top\in\mathbb{R}^p$ is a parameter vector for fixed effects, ${\bf Z}=({\bm z}_1^\top,\dots,{\bm z}_n^\top)^\top\in\mathbb{R}^{n\times d}$ with ${\bm z}_{i}=(z_{i1},\dots,z_{id})^\top\in\mathbb{R}^d$ is a design matrix for random effects, and ${\bm\gamma}=(\gamma_1,\dots,\gamma_d)^\top\in\mathbb{R}^{d}$ is a random effects vector. Both ${\bf X}$ and ${\bf Z}$ are full rank, implying that they satisfy ${\rm rank}({\bf X})=p$ and ${\rm rank}({\bf Z})=d$ in~\eqref{eq:generalized linear mixed model}. A prior distribution for ${\bm\gamma}$ is needed to specify the GLMM. A multivariate normal distribution is often used, such that the prior density for ${\bm\gamma}$ can be expressed as $\pi({\bm\gamma})=\pi_{\bm\omega}({\bm\gamma})=\varphi({\bm\gamma};{\bm 0},{\bf D}_{\bm\omega})$ with
\begin{equation}
\label{eq:multivariate normal density}
\varphi({\bm t};{\bm \nu},{\bm\Sigma})={1\over (2\pi)^{k/2}|\det({\bm\Sigma})|^{1/2}}e^{-{1\over 2}({\bm t}-{\bm\nu})^\top{\bm\Sigma}^{-1}({\bm t}-{\bm\nu})}
\end{equation}
for the PDF of ${\mathcal N}({\bm\nu},{\bm\Sigma})$, where ${\bm\nu}$ is a $k$-dimensional mean vector and ${\bm\Sigma}$ is a $k\times k$-dimensional covariance matrix. In practice, longitudinal data or spatial models often specify ${\bf D}$. 

The GLMM jointly defined by~\eqref{eq:exponential family distribution},~\eqref{eq:generalized linear mixed model}, and~\eqref{eq:distribution of random effects} is a special case of Bayesian Hierarchical Models (BHMs) with two hierarchical levels. The first level, specified by~\eqref{eq:exponential family distribution} and \eqref{eq:generalized linear mixed model}, provides $f({\bm y}|{\bm\gamma})$ the conditional distribution of ${\bm y}|{\bm\gamma}$. The second level, specified by \eqref{eq:distribution of random effects}, provides ${\bm\gamma}\sim\pi({\bm\gamma})$ the prior distribution for the random effects. The choice of the multivariate normal distribution for ${\bm\gamma}$ is convenient for modeling dependencies between the random effects. 

The exponential family distribution given by~\eqref{eq:exponential family distribution} satisfies ${\bm\mu}=(\mu_1,\dots,\mu_n)^\top=b'({\bm\eta})={\rm E}({\bm y}|{\bm\eta})$ and ${\rm cov}({\bm y}|{\bm\eta})={\rm diag}\{b''({\bm\eta})\}$, where $\mu_i=b'(\eta_i)={\rm E}(y_i|{\bm\eta})$ and ${\rm var}(y_i|{\bm\eta})=b''(\eta_i)$. Under~\eqref{eq:generalized linear mixed model} and~\eqref{eq:distribution of random effects}, the GLMM can be expressed as ${\bm y}|{\bm\gamma}\sim f({\bm y}|{\bm\gamma})$ and ${\bm\gamma}\sim\pi({\bm\gamma})$. Because ${\bm\gamma}$ is unobserved, the likelihood function (or the marginal PMF/PDF) of the GLMM is derived by integrating out ${\bm\gamma}$ from $L({\bm\beta},{\bm\omega}|{\bm\gamma})=f({\bm y}|{\bm\gamma})\pi({\bm\gamma})$ as
\begin{equation}
\label{eq:likelihood function marginal}
L({\bm\beta},{\bm\omega})=\bar f({\bm y})=\bar f_{{\bm\beta}{\bm\omega}}({\bm y})=\int_{\mathbb{R}^d}L({\bm\beta},{\bm\delta}|{\bm\gamma})d{\bm\gamma}.
\end{equation}

The MLE of the parameters and the hyperparameters is defined as
\begin{equation}
\label{eq:MLE of parameters and hyperparameters}
(\hat{\bm\beta}^\top,\hat{\bm\omega}^\top)^\top=\mathop{\arg\!\max}_{{\bm\beta},{\bm\omega}} \ell({\bm\beta},{\bm\omega}),
\end{equation}
where $\ell({\bm\beta},{\bm\omega})=\log L({\bm\beta},{\bm\omega})$ is the loglikelihood function. Because~\eqref{eq:MLE of parameters and hyperparameters} cannot be analytically solved, the MLE is investigated based on a root of the likelihood equation (i.e., the score equation) as
\begin{equation}
\label{eq:MLE for score equation likelihood}
\dot\ell(\check{\bm\beta},\check{\bm\omega})={\bm 0},
\end{equation}
where $\dot\ell({\bm\beta},{\bm\omega})=(\dot\ell_{\bm\beta}^\top({\bm\beta},{\bm\omega}),\dot\ell_{\bm\omega}^\top({\bm\beta},{\bm\omega}))^\top$ is the gradient of $\ell({\bm\beta},{\bm\omega})$ with respect to ${\bm\beta}$ and ${\bm\omega}$, $\dot\ell_{\bm\beta}({\bm\beta},{\bm\omega})=(\partial\ell({\bm\beta},{\bm\omega})/\partial\beta_0,\dots,\partial\ell({\bm\beta},{\bm\omega})/\partial\beta_{p-1})^\top$ is the gradient of  $\ell({\bm\beta},{\bm\omega})$ with respect to ${\bm\beta}$, and $\dot\ell_{\bm\omega}({\bm\beta},{\bm\omega})=(\partial  \ell({\bm\beta},{\bm\omega})/\partial\omega_1,\dots,\partial\ell({\bm\beta},{\bm\omega})/\partial\omega_{r})^\top$ is the gradient vector of  $\ell({\bm\beta},{\bm\omega})$ with respect to ${\bm\omega}$. The MLE belongs to the solution set of~\eqref{eq:MLE for score equation likelihood} as
\begin{equation}
\label{eq:solution set of MLE for score equation likelihood}
{\mathcal A}=\{({\bm\beta}^\top,{\bm\omega}^\top)^\top: \dot\ell({\bm\beta},{\bm\omega})={\bm 0}\}.
\end{equation}
There is $(\check{\bm\beta}^\top,\check{\bm\omega}^\top)^\top=(\hat{\bm\beta},\hat{\bm\omega}^\top)^\top$ if $|{\mathcal A}|=1$; or $(\check{\bm\beta}^\top,\check{\bm\omega}^\top)^\top$ and $(\hat{\bm\beta},\hat{\bm\omega}^\top)^\top$ may be different otherwise. 

Computing a solution to~\eqref{eq:MLE for score equation likelihood} is the first step to the exact MLE problem defined by~\eqref{eq:MLE of parameters and hyperparameters}. The solution may not be a global optimizer. We need to evaluate~\eqref{eq:MLE of parameters and hyperparameters}. We show that the prediction-maximization (PM) algorithm proposed by~\cite{zhang2021,zhang2023} can be used to compute a solution to~\eqref{eq:MLE for score equation likelihood}. The PM algorithm has a P-step for predicting the random effects and an M-step for maximizing the objective function. Under the framework of the PM algorithm, we construct a sequence of objective functions such that they can induce an exact solution to~\eqref{eq:MLE for score equation likelihood}. We solve the exact MLE problem although the likelihood function given by~\eqref{eq:likelihood function marginal} keeps unsolved. We present our method in Section~\ref{sec:main result}.


The integral on the right-hand size of~\eqref{eq:likelihood function marginal} is intractable if $f({\bm y}|{\bm\gamma})$ given by~\eqref{eq:exponential family distribution} is not normal. This is considered the main difficulty in the MLE for GLMMs when the response does not follow normal. Previous methods are mostly developed under the framework of the Laplace approximation (LA) approach \cite[Section 3.3]{barndoeff-nielsen1989}. Based on a second-order Taylor expansion, the LA provides an approximation of the right-hand side of~\eqref{eq:likelihood function marginal} as
\begin{equation}
\label{eq:laplace approximation formula}
 L({\bm\beta},{\bm\omega})
\approx  L_{LA}({\bm\beta},{\bm\omega})={(2\pi)^{d/2} L({\bm\beta},{\bm\omega}|\breve{\bm\gamma}_{{\bm\beta}{\bm\omega}})\over |-\det\{\Delta \log[L({\bm\beta},{\bm\omega}|\breve{\bm\gamma}_{{\bm\beta}{\bm\omega}})]\}|^{1/2}},
\end{equation}
where $\Delta \log[L({\bm\beta},{\bm\omega}|\breve{\bm\gamma}_{{\bm\beta}{\bm\delta}})]$ is the Hessian matrix of $\log[L({\bm\beta},{\bm\omega}|{\bm\gamma})]$ with respect to ${\bm\gamma}$ at 
\begin{equation}
\label{eq:maximizer used in the Laplace approximation}
\breve\gamma_{{\bm\beta}{\bm\omega}}=\arg\!\max_{\bm\gamma} \log[L({\bm\beta},{\bm\omega}|{\bm\gamma})].
\end{equation}
To apply the LA approach, one has to compute the gradient vector and Hessian matrix of $ L_{LA}({\bm\beta},{\bm\omega})$ concerning ${\bm\beta}$ and ${\bm\omega}$. The calculation of $\breve{\bm\gamma}_{{\bm\beta}{\bm\omega}}$ is also needed. The optimization procedure is often complicated and unreliable when $d$ is large. The Monte Carlo approach may be used to address this issue. An example is a method incorporated in the \textsf{PrevMap} package of \textsf{R}~\cite{giorgi2017}.

The LA is treated as a fundamental technique in computing the approximate MLEs of ${\bm\beta}$ and ${\bm\omega}$. Many methods have been developed under the LA. Examples include the penalized quasi-likelihood (PQL)~\cite{breslow1993} and the integrated nested Laplace approximation (INLA)~\cite{rue2009} methods. It has been pointed out that the second-order LA given by~\eqref{eq:laplace approximation formula} can cause serious bias and a bias correction is needed~\cite{breslow1995}. To make the approximation more accurate, the higher-order LA approach is proposed~\cite{evangelou2011}. The approach uses the expressions of the third and fourth partial derivatives of $\log[L({\bm\beta},{\bm\omega}|{\bm\gamma})]$ with respect to ${\bm\gamma}$ at $\breve{\bm\gamma}_{{\bm\beta}{\bm\omega}}$, which are present on the right-hand side of~\eqref{eq:laplace approximation formula}. The expressions of the third and fourth moments are complicated. Many of those have been incorporated into software packages. Examples include the \textsf{spaMM} package of \textsf{R}~\cite{rousset2021}. 

Previous approximate MLE methods are mostly developed under~\eqref{eq:laplace approximation formula} or another formulation of $L_{LA}({\bm\beta},{\bm\omega})$ provided by the higher-order LA. They optimize the approximate likelihood $ L_{LA}({\bm\beta},{\bm\omega})$ but not the exact likelihood $L({\bm\beta},{\bm\omega})$. A solution given by $(\breve{\bm\beta}_{LA}^\top,\breve{\bm\omega}_{LA}^\top)^\top=\arg\!\max_{{\bm\beta},{\bm\omega}} L_{LA}({\bm\beta},{\bm\omega})$ does not satisfies~\eqref{eq:MLE for score equation likelihood}, while the solution of our method satisfies~\eqref{eq:MLE for score equation likelihood}.


The {\rm EM} and the MCMC approaches are also popular in computing the approximate MLE of ${\bm\beta}$ and ${\bm\omega}$. The EM is carried out by forming the loglikelihood of the complete data. It calculates the expectation concerning the conditional distribution of ${\bm\gamma}$ given ${\bm y}$ in the E-step and then maximizes it concerning ${\bm\beta}$ and ${\bm\omega}$ in the M-step. Because the E-step cannot be computed in closed forms, a Monte Carlo approach often carries the EM algorithm for GLMMs, leading to an MCEM algorithm~\cite{mcculloch1997}. The MCMC is carried out by a Metropolis-Hastings algorithm, which generates a Markov chain sequence of values that eventually stabilizes at the posterior distribution. Many MCMC methods have been proposed and incorporated into software packages. Examples include the \textsf{geoRglm}~\cite{christensen2002} and the \textsf{geoCount}~\cite{jing2015} packages of \textsf{R}. However, a previous study shows that the PQL and the LA outperform the MCMC methods in general~\cite{zhang2024}.

The intractable integral problem is connected with Louville's Theorem in differential algebra, a field of mathematics. The corresponding mathematical problem is called integration in closed forms, meaning that an integral or antiderivative can be expressed as an elementary function or closed form. French mathematician Joseph Liouville first studied this problem in 1833~\cite{liouville1933a,liouville1933b,liouville1933c}. Liouville's work received little attention until the 1940s when Ritt~\cite{ritt1948} developed a field called {\it differential algebra} in mathematics. A growing interest following Liouville's work has appeared due to advances in computer languages of symbolic mathematical computation. The corresponding mathematical research has evolved significantly since the work of~\cite{risch1969,rosenlicht1972}. The domain of the integral on the right-hand sides of~\eqref{eq:likelihood function marginal} is the entire $\mathbb{R}^d$. It is also treated as a special integral. The special integral techniques can be used, leading to our method. 

The article is organized as follows. In Section~\ref{sec:liouville's theorem in mathematics}, we review Liouville's Theorem in mathematics. In Section~\ref{sec:main result}, we present our main result. In Section~\ref{sec:test statistics}, we formulate a few test statistics for the significance of parameters and hyperparameters. In Section~\ref{sec:simulation}, we evaluate the performance of our method by Monte Carlo simulations. In Section~\ref{sec:application}, we apply our method to a real-world dataset. In Section~\ref{sec:conclusion}, we conclude the article. We put all proofs in the Appendix.

\section{Liouville's Theorem in Mathematics}
\label{sec:liouville's theorem in mathematics}

When solving for an indefinite integral of a given elementary function $f(x)$, the fundamental problem of elementary calculus is to answer whether the antiderivative $F(x)=\int_a^x f(t)dt$ can be expressed as an elementary function. In mathematics, an elementary function is a function of a single variable that can be expressed as sums, products, roots, and compositions of constants, rational powers, exponential, logarithms, trigonometric, inverse trigonometric, hyperbolic, inverse hyperbolic functions as well as their finite additions, subtractions, multiplications, divisions, roots, and compositions~\cite[Chapter 12]{geddes1992}. The problem of integration in closed form or the problem of integration in finite terms is well-known in differential algebra. The ideal case is that there is an elementary function identical to $F(x)$ for any $x$ in its domain. Liouville's Theorem states that the ideal case may not be achieved due to the nonexistence of elementary $F(x)$. 

Given that $F(x)$ is nonelementary, $F(x)$ may still be solved in special cases. A typical example is $f(x)=x^{-1}\sin{x}$, where $F(x)=\int_0^x t^{-1}\sin{t}dt$ nonelementary~\cite{williams1993}. As a special integral, $F(\infty)=\int_0^\infty t^{-1}\sin{t}dt$ can be solved via $F(\infty)=\lim_{\alpha\rightarrow 0^+} \int_0^\infty e^{-\alpha{t}}t^{-1}\sin{t}dt$. Let $h(\alpha)=\int_0^\infty e^{-\alpha{t}} t^{-1}\sin{t} dt$. By $h'(\alpha)=-\int_0^\infty e^{-\alpha{t}} \sin{t} dt=-1/(1+\alpha^2)$ and $h(\infty)=0$, there is $h(\alpha)= \int_\alpha^\infty h'(t)dt={\pi/ 2}-\arctan\alpha$. The definite integral is
\begin{equation}
\label{eq:sin(x)/x special integral}
F(\infty)=\int_0^\infty {\sin{t}\over t}dt={\pi\over 2},
\end{equation}
although the indefinite integral 
\begin{equation}
\label{eq:sin(x)/x antiderivatives}
F(x)=\int_0^x {\sin(t)\over t}dt 
\end{equation}
is nonelementary and remains unsolved.

In mathematics, the definite integral given by~\eqref{eq:sin(x)/x special integral} is classified as a special integral (another term is a specific definite integral). The indefinite integral given by~\eqref{eq:sin(x)/x antiderivatives} is classified as a nonelementary antiderivative, also called an intractable integral. The lower or upper bounds of the intractable integrals can be arbitrary. A special integral is derived if the lower and upper bounds  take values in $\{-\infty,0,\infty\}$. Given that an antiderivative is intractable, the corresponding special integral can only be studied specifically. The method for~\eqref{eq:sin(x)/x special integral} may not work if another special integral is considered.  

\section{Main Result}
\label{sec:main result}

The right-hand side of~\eqref{eq:likelihood function marginal} can be treated as an intractable or special integral. The intractable integral approach examines the corresponding indefinite integral with the domain to be arbitrary. The right-hand side of~\eqref{eq:likelihood function marginal} is obtained if the domain goes to the entire $\mathbb{R}^d$. In mathematics, an intractable integral is often evaluated by a Taylor series. In many cases, the coefficients are also nonelementary even if the Taylor series converges. They are still difficult to compute. The special integral approach does not change the domain. It devises a specific tool to handle the special integral.  We adopt the special integral approach in our research. We study~\eqref{eq:MLE for score equation likelihood} and then~\eqref{eq:MLE of parameters and hyperparameters}. 


The $j$th component of $\dot\ell_{\bm\beta}({\bm\beta},{\bm\omega})$ is
\begin{equation}
\label{eq:partial derivative of the loglikelihood beta}
{\partial\ell({\bm\beta},{\bm\omega})\over\partial\beta_j}= \int_{\mathbb{R}^d} {\bm x}_j^\top[{\bm y}-b'({\bm\eta}_{{\bm\beta}{\bm\gamma}})] q_{{\bm\beta}{\omega}}({\bm\gamma}|{\bm y})d{\bm\gamma}
\end{equation}
for $j=0,\dots,p-1$, where
\begin{equation}
\label{eq:posterior density of gamma given the response}
q_{{\bm\beta}{\bm\omega}}({\bm\gamma}|{\bm y})={ f_{\bm\beta}({\bm y}|{\bm\gamma})\pi_{\bm\omega}({\bm\gamma})\over \bar f_{{\bm\beta}{\bm\omega}}({\bm y})}
\end{equation}
is the posterior density of ${\bm\gamma}$ given ${\bm y}$. The $j$th component of $\dot\ell_{\bm\omega}({\bm\beta},{\bm\omega})$ is 
\begin{equation}
\label{eq:partial derivative of the loglikelihood delta}
{\partial\ell({\bm\beta},{\bm\omega})\over\partial\omega_j}= \int_{\mathbb{R}^d} \left[-{1\over 2}{\rm tr}\left({\bf D}_{\bm\omega}^{-1}{\partial{\bf D}_{\bm\omega}\over\partial\omega_j}\right)+{1\over 2}{\bm\gamma}^\top{\bf D}_{\bm\omega}^{-1}{\partial{\bf D}_{\bm\omega}\over\partial\omega_j}{\bf D}_{\bm\omega}^{-1}{\bm\gamma}\right] q_{{\bm\beta}{\bm\omega}}({\bm\gamma}|{\bm y})d{\bm\gamma}
\end{equation}
for $j=1,\dots,r$. 

We use~\eqref{eq:partial derivative of the loglikelihood beta} and~\eqref{eq:partial derivative of the loglikelihood delta} to construct an objective function for computing a solution of~\eqref{eq:MLE for score equation likelihood}. We define an initial objective function as
\begin{equation}
\label{eq:constuction of initial working objective function for MLE}
\eqalign{
h_{{\bm\beta}{\bm\omega}}({\bm\alpha},{\bm\delta})
=&\int_{\mathbb{R}^d}\log[\varphi(\tilde{\bm y}_{{\bm\beta}{\bm\gamma}}; {\bf X}{\bm\beta}+{\bf X}{\bm\alpha}+{\bf Z}{\bm\gamma},{\bf W}_{{\bm\beta}{\bm\gamma}}^{-1})\varphi({\bm\gamma};{\bm 0},{\bf D}_{{\bm\omega}+{\bm\delta}})]q_{{\bm\beta}{\bm\omega}}({\bm\gamma}|{\bm y})d{\bm\gamma},
}\end{equation}
where $\tilde{\bm y}_{{\bm\beta}{\bm\gamma}}={\bm\eta}_{{\bm\beta}{\bm\gamma}}+[{\bm y}-b'({\bm\eta}_{{\bm\beta}{\bm\gamma}})]/b''({\bm\eta}_{{\bm\beta}{\bm\gamma}})$ and ${\bf W}_{{\bm\beta}{\bm\gamma}}={\rm diag}[b''({\bm\eta}_{{\bm\beta}{\bm\gamma}})]$. 

Let $\dot h_{{\bm\beta}{\bm\omega}}({\bm\alpha},{\bm\delta})=(\dot h_{{\bm\beta}{\bm\delta},{\bm\alpha}}^\top({\bm\alpha},{\bm\delta}),\dot h_{{\bm\beta}{\bm\delta},{\bm\delta}}^\top({\bm\alpha},{\bm\delta}))^\top$ be the gradient of $h_{{\bm\beta}{\bm\omega}}({\bm\alpha},{\bm\delta})$ with respect to ${\bm\alpha}$ and ${\bm\delta}$. The $j$th component of $\dot h_{{\bm\beta}{\bm\delta},{\bm\alpha}}({\bm\alpha},{\bm\delta})$ is
\begin{equation}
\label{eq:partial derivative of initial objective function beta for the MLE}
\eqalign{
{\partial h_{{\bm\beta}{\bm\omega}}({\bm\alpha},{\bm\delta})\over\partial\alpha_j}=&\int_{\mathbb{R}^d}  {\bm x}_j^\top {\bf W}_{{\bm\beta}{\bm\gamma}}(\tilde{\bm y}_{{\bm\beta}{\bm\gamma}}-{\bf X}{\bm\beta}-{\bf X}{\bm\alpha}-{\bf Z}{\bm\gamma})q_{{\bm\beta}{\bm\omega}}({\bm\gamma}|{\bm y})d{\bm\gamma},
}
\end{equation}
for $j=0,\dots,p-1$. The $j$th component of $\dot h_{{\bm\beta}{\bm\omega},{\bm\delta}}({\bm\alpha},{\bm\delta})$ is
\begin{equation}
\label{eq:partial derivative of initial objective function omega for the MLE}
\eqalign{
{\partial h_{{\bm\beta}{\bm\omega}}({\bm\alpha},{\bm\delta})\over\partial\delta_j}=&\int_{\mathbb{R}^d} \left[-{1\over 2}{\rm tr}\left({\bf D}_{{\bm\omega}+{\bm\delta}}^{-1}{\partial  {\bf D}_{{\bm\omega}+{\bm\delta}}\over\partial\delta_j}\right)+{1\over 2}{\bm\gamma}^\top{\bf D}_{{\bm\omega}+{\bm\delta}}^{-1}{\partial{\bf D}_{{\bm\omega}+{\bm\delta}}\over\partial\delta_j} {\bf D}_{{\bm\omega}+{\bm\delta}}^{-1}{\bm\gamma}  \right] q_{{\bm\beta}{\bm\omega}}({\bm\gamma}|{\bm y})d{\bm\gamma}
}
\end{equation}
for $j=1,\dots,r$. We compare~\eqref{eq:partial derivative of the loglikelihood beta} with~\eqref{eq:partial derivative of initial objective function beta for the MLE}, and~\eqref{eq:partial derivative of the loglikelihood delta} with~\eqref{eq:partial derivative of initial objective function omega for the MLE}. We obtain
\begin{equation}
\label{eq:equality for the gradient of the loglikelihood and the working loglikelihood}
\dot h_{{\bm\beta}{\bm\omega}}({\bm 0},{\bm 0})=\dot\ell({\bm\beta},{\bm\omega}),
\end{equation}
implying that~\eqref{eq:solution set of MLE for score equation likelihood} becomes
\begin{equation}
\label{eq:identity for gradients of beta of the loglikelihood and initial working objective function of MLE}
{\cal S}=\{({\bm\beta}^\top,{\bm\omega}^\top)^\top: \dot h_{{\bm\beta}{\bm\omega}}({\bm 0},{\bm 0})={\bm 0}\}.
\end{equation}
Because $h_{{\bm\beta}{\bm\omega}}({\bm\alpha},{\bm\delta})$ given by~\eqref{eq:constuction of initial working objective function for MLE} is constructed based on normal densities, we avoid the computational difficulty caused by the intractable integral on the right-hand side of~\eqref{eq:likelihood function marginal}.

We compute the Hessian matrix
\begin{equation}
\label{eq:hessian matrix of the initial working objective function for MLE}
\ddot h_{{\bm\beta}{\bm\omega}}({\bm\alpha},{\bm\delta})=\left(\begin{array}{cc} \ddot h_{{\bm\beta}{\bm\omega},{\bm\alpha}{\bm\alpha}}({\bm\alpha},{\bm\beta}) & \ddot h_{{\bm\beta}{\bm\omega},{\bm\alpha}{\bm\delta}}({\bm\alpha},{\bm\beta}) \cr  \ddot h_{{\bm\beta}{\bm\omega},{\bm\delta}{\bm\alpha}}({\bm\alpha},{\bm\beta}) &  \ddot h_{{\bm\beta}{\bm\omega},{\bm\delta}{\bm\delta}}({\bm\alpha},{\bm\beta})\end{array}\right)
\end{equation}
of $h_{{\bm\beta}{\bm\omega}}({\bm\alpha},{\bm\delta})$. We have $\ddot h_{{\bm\beta}{\bm\omega},{\bm\delta}{\bm\alpha}}^\top ({\bm\alpha},{\bm\beta})=\ddot h_{{\bm\beta}{\bm\omega},{\bm\alpha}{\bm\delta}}({\bm\alpha},{\bm\beta})={\bm 0}$, 
\begin{equation}
\label{eq:second-order partial derivative of our objective function alpha for the MLE}
\eqalign{
\ddot h_{{\bm\beta}{\bm\omega},{\bm\alpha}{\bm\alpha}}({\bm\alpha},{\bm\delta})=&-\int_{\mathbb{R}^d} {\bf X}^\top b''({\bm\eta}_{{\bm\beta}\bm\gamma}){\bf X}  q_{{\bm\beta}{\bm\omega}}({\bm\gamma}|{\bm y})d{\bm\gamma}, 
}
\end{equation}
and the $(j_1,j_2)$th component of $\ddot h_{{\bm\beta}{\bm\omega},{\bm\delta}{\bm\delta}}({\bm\alpha},{\bm\beta})$ for $j_1,j_2=1,\dots,r$ as
\begin{equation}
\label{eq:second-order partial derivative of our objective function delta for the MLE}
\eqalign{
{\partial h_{{\bm\beta}{\bm\omega}}({\bm\alpha},{\bm\delta})\over\partial\delta_{j_1}\partial\delta_{j_2}}=&\int_{\mathbb{R}^d} \left[-{1\over 2}{\rm tr}\left({\bf D}_{{\bm\omega}+{\bm\delta}}^{-1}{\partial  {\bf D}_{{\bm\omega}+{\bm\delta}}\over\partial\delta_j}\right) +{1\over 2}{\bm\gamma}^\top{\bf D}_{{\bm\omega}+{\bm\delta}}^{-1}{\partial{\bf D}_{{\bm\omega}+{\bm\delta}}\over\partial\delta_{j_1}\partial\delta_{j_2}} {\bf D}_{{\bm\omega}+{\bm\delta}}^{-1}{\bm\gamma} \right. \cr
&+{1\over 2}{\rm tr}\left({\bf D}_{{\bm\omega}+{\bm\delta}}^{-1}{\partial  {\bf D}_{{\bm\omega}+{\bm\delta}}\over\partial\delta_{j_1}}{\bf D}_{{\bm\omega}+{\bm\delta}}^{-1}{\partial  {\bf D}_{{\bm\omega}+{\bm\delta}}\over\partial\delta_{j_2}}\right)\cr
&\left. -{\bm\gamma}^\top{\bf D}_{{\bm\omega}+{\bm\delta}}^{-1}{\partial{\bf D}_{{\bm\omega}+{\bm\delta}}\over\partial\delta_{j_1}}{\bf D}_{{\bm\omega}+{\bm\delta}}^{-1}{\partial{\bf D}_{{\bm\omega}+{\bm\delta}}\over\partial\delta_{j_2}} {\bf D}_{{\bm\omega}+{\bm\delta}}^{-1}{\bm\gamma}  \right] q_{{\bm\beta}{\bm\omega}}({\bm\gamma}|{\bm y})d{\bm\gamma}.
}
\end{equation}

Because $\ddot h_{{\bm\beta}{\bm\omega},{\bm\alpha}{\bm\alpha}}({\bm\alpha},{\bm\delta})$ is negative definite, the solution ${\bm\alpha}_0$ to $\dot h_{{\bm\beta}{\bm\omega},{\bm\alpha}}({\bm\alpha},{\bm\delta})={\bm 0}$ is unique. Because $\dot h_{{\bm\beta}{\bm\omega},{\bm\alpha}}({\bm\alpha},{\bm\delta})=\dot h_{{\bm\beta}{\bm\omega},{\bm\alpha}}({\bm\alpha},{\bm 0})$ for all ${\bm\delta}$, ${\bm\alpha}_0$ does not depend on ${\bm\delta}$. By the Bolzano–Poincar\'e–Miranda theorem in mathematics~\cite[e.g.]{vrahatis2016}, there exists a unique $\hat{\bm\gamma}_{{\bm\beta}{\bm\omega}}$ that does not depend on ${\bm\gamma}$ and ${\bm\delta}$, such that we can use it to replace ${\bm\gamma}$ in $\tilde{\bm y}_{{\bm\beta}{\bm\gamma}}$ and ${\bf W}_{{\bm\delta}{\bm\gamma}}$, leading to the usage of $\tilde{\bm y}_{{\bm\beta}\hat{\bm\gamma}_{{\bm\beta}{\bm\omega}}}$ and ${\bf W}_{{\bm\delta}\hat{\bm\gamma}_{{\bm\beta}{\bm\omega}}}$ in~\eqref{eq:partial derivative of initial objective function beta for the MLE}, respectively. It satisfies
\begin{equation}
\label{eq:partial derivative of initial objective function beta for MLE BPM}
\dot h_{{\bm\beta}{\bm\omega},{\bm\alpha}}({\bm\alpha}_0,{\bm\delta})=\int_{\mathbb{R}^d}{\bf X}^\top{\bf W}_{{\bm\beta}\hat{\bm\gamma}_{{\bm\beta}{\bm\omega}}}(\tilde{\bm y}_{{\bm\beta}\hat{\bm\gamma}_{{\bm\beta}{\bm\omega}}}-{\bf X}{\bm\beta}-{\bf X}{\bm\alpha}_0-{\bf Z}{\bm\gamma})q_{{\bm\beta}{\bm\omega}}({\bm\gamma}|{\bm y})d{\bm\gamma}={\bm 0}
\end{equation}
and $\dot h_{{\bm\beta}{\bm\omega},{\bm\alpha}}({\bm 0},{\bm\delta})={\bm 0}$ for any $({\bm\beta}^\top,{\bm\omega}^\top)^\top\in{\mathcal A}$, leading to a modification of~\eqref{eq:constuction of initial working objective function for MLE} as
\begin{equation}
\label{eq:modified working objective function for MLE}
\eqalign{
\tilde h_{{\bm\beta}{\bm\omega}}({\bm\alpha},{\bm\delta})
=&\int_{\mathbb{R}^d}\log[\varphi(\tilde{\bm y}_{{\bm\beta}\hat{\bm\gamma}_{{\bm\beta}{\bm\omega}}}; {\bf X}{\bm\beta}+{\bf X}{\bm\alpha}+{\bf Z}{\bm\gamma},{\bf W}_{{\bm\beta}\hat{\bm\gamma}_{{\bm\beta}{\bm\omega}}}^{-1}) \varphi({\bm\gamma};{\bm 0},{\bf D}_{{\bm\omega}+{\bm\delta}})]q_{{\bm\beta}{\bm\omega}}({\bm\gamma}|{\bm y})d{\bm\gamma}.\cr
}
\end{equation}

\begin{lem}
\label{lem:normal density posterior identity}
$\tilde h_{{\bm\beta}{\bm\omega}}({\bm\alpha},{\bm\delta})$ given by~\eqref{eq:modified working objective function for MLE} satisfies $\hat{\bm\gamma}_{{\bm\beta}{\bm\omega}}={\bm v}_{{\bm\beta}{\bm\omega},{\bm 0}{\bm 0}}$ and 
\begin{equation}
\label{eq:normal density identity for initial modified objective function}
\eqalign{
&\varphi(\tilde{\bm y}_{{\bm\beta}\hat{\bm\gamma}_{{\bm\beta}{\bm\omega}}};{\bf X}{\bm\beta}+{\bf X}{\bm\alpha}+{\bf Z}{\bm\gamma},{\bf W}_{{\bm\beta}\hat{\bm\gamma}_{{\bm\beta}{\bm\omega}}}^{-1})\varphi({\bm \gamma};{\bf 0},{\bf D}_{{\bm\omega}+{\bm\delta}})\cr
=&\varphi({\bm\gamma}; {\bm v}_{{\bm\beta}{\bm\omega},{\bm\delta}},{\bf V}_{{\bm\beta}{\bm\omega},{\bm\delta}})\varphi(\tilde{\bm y}_{{\bm\beta}\hat{\bm\gamma}_{{\bm\beta}{\bm\omega}}}; {\bf X}{\bm\beta}+{\bf X}{\bm\alpha},{\bf R}_{{\bm\beta}{\bm\omega},{\bm\delta}}),\cr
}
\end{equation}
where ${\bm v}_{{\bm\beta}{\bm\omega},{\bm\alpha}{\bm\delta}}={\bf D}_{{\bm\omega}+{\bm\delta}}{\bf Z}^\top{\bf R}_{{\bm\beta}{\bm\omega},{\bm\delta}}^{-1}(\tilde{\bm y}_{{\bm\beta}\hat{\bm\gamma}_{{\bm\beta}{\bm\omega}}}-{\bf X}{\bm\beta}-{\bf X}{\bm\alpha})$, ${\bf V}_{{\bm\beta}{\bm\omega},{\bm\delta}}={\bf D}_{{\bm\omega}+{\bm\delta}}-{\bf D}_{{\bm\omega}+{\bm\delta}}{\bf Z}^\top{\bf R}_{{\bm\beta}{\bm\omega},{\bm\delta}}^{-1}{\bf Z}{\bf D}_{{\bm\omega}+{\bm\delta}}$, and ${\bf R}_{{\bm\beta}{\bm\omega},{\bm\delta}}={\bf W}_{{\bm\beta}\hat{\bm\gamma}_{{\bm\beta}{\bm\omega}}}^{-1}+{\bf Z}{\bf D}_{{\bm\omega}+{\bm\delta}}{\bf Z}^\top$.
\end{lem}

Lemma~\ref{lem:normal density posterior identity} indicates that we can replace $h_{{\bm\beta}{\bm\omega}}({\bm\alpha},{\bm\delta})$ with
\begin{equation}
\label{eq:modified working objective function for MLE for computing}
\eqalign{
\psi_{{\bm\beta}{\bm\omega}}({\bm\alpha},{\bm\delta})=&\log[\varphi(\tilde{\bm y}_{{\bm\beta}\hat{\bm\gamma}_{{\bm\beta}{\bm\omega}}}; {\bf X}{\bm\beta}+{\bf X}{\bm\alpha},{\bf R}_{{\bm\beta}{\bm\omega},{\bm\delta}})] 
}\end{equation}
in the optimization procedure. We show that it is appropriate.

\begin{thm}
\label{thm:gradient property of modified working objective function for MLE}
(Main Theorem) $\dot\ell({\bm\beta},{\bm\omega})={\bm 0}\Leftrightarrow  \dot h_{{\bm\beta}{\bm\omega}}({\bm 0},{\bm 0})={\bm 0}\Leftrightarrow\dot\psi_{{\bm\beta}{\bm\omega}}({\bm 0},{\bm 0})={\bm 0}$ and ${\mathcal A}=\{({\bm\beta}^\top,{\bm\omega}^\top)^\top: \dot\psi_{{\bm\beta}{\bm\omega}}({\bm 0},{\bm 0})={\bm 0}\}$.
\end{thm}

 We compute $\dot\psi_{{\bm\beta}{\bm\omega}}({\bm\alpha},{\bm\delta})=(\dot\psi_{{\bm\beta}{\bm\omega},{\bm\alpha}}^\top({\bm\alpha},{\bm\delta}),\dot\psi_{{\bm\beta}{\bm\omega},{\bm\delta}}^\top({\bm\alpha},{\bm\delta}))^\top$. The gradient vector $\dot\psi_{{\bm\beta}{\bm\omega},{\bm\alpha}}({\bm\alpha},{\bm\delta})$ of $\psi_{{\bm\beta}{\bm\omega}}({\bm\alpha},{\bm\delta})$ with respect to ${\bm\alpha}$ is
\begin{equation}
\label{eq:gradient vector of modified working objective function for MLE for computing for alpha}
\eqalign{
\dot\psi_{{\bm\beta}{\bm\omega},{\bm\alpha}}({\bm\alpha},{\bm\delta})=&{\bf X}^\top{\bf R}_{{\bm\beta}{\bm\omega},{\bm\delta}}^{-1}(\tilde{\bm y}_{{\bm\beta}\hat{\bm\eta}_{{\bm\beta}{\bm\omega}}}-{\bf X}{\bm\beta}-{\bf X}{\bm\alpha}).
}\end{equation}
The $j$th component of the gradient vector $\dot\psi_{{\bm\beta}{\bm\omega},{\bm\omega}}({\bm\alpha},{\bm\delta})$ of $\psi_{{\bm\beta}{\bm\omega}}({\bm\alpha},{\bm\delta})$ with respect to ${\bm\delta}$ is
\begin{equation}
\label{eq:gradient vector of modified working objective function for MLE for computing for delta}
\eqalign{
{\partial\psi_{{\bm\beta}{\bm\omega}}({\bm\alpha},{\bm\delta})\over\partial\delta_j}=&-{1\over 2}{\rm tr}\left( {\bf R}_{{\bm\beta}{\bm\omega},{\bm\delta}}^{-1}{\partial {\bf R}_{{\bm\beta}{\bm\omega},{\bm\delta}}\over\partial\delta_j}\right)+{1\over 2}(\tilde{\bm y}_{{\bm\beta}\hat{\bm\eta}_{{\bm\beta}{\bm\omega}}}-{\bf X}{\bm\beta}-{\bf X}{\bm\alpha})^\top {\bf R}_{{\bm\beta}{\bm\omega},{\bm\delta}}^{-1}\cr
&\hspace{2cm}{\partial {\bf R}_{{\bm\beta}{\bm\omega},{\bm\delta}}\over\partial\delta_j}{\bf R}_{{\bm\beta}{\bm\omega},{\bm\delta}}^{-1} (\tilde{\bm y}_{{\bm\beta}\hat{\bm\eta}_{{\bm\beta}{\bm\omega}}}-{\bf X}{\bm\beta}-{\bf X}{\bm\alpha}) 
}\end{equation}
for $j=1,\dots,r$. We compute the Hessian matrix 
\begin{equation}
\label{eq:hessian matrix modified working objective function for MLE}
\ddot\psi_{{\bm\beta}{\bm\omega}}({\bm\alpha},{\bm\delta})=\left(\begin{array}{cc}  \ddot\psi_{{\bm\beta}{\bm\omega},{\bm\alpha}{\bm\alpha}}({\bm\alpha},{\bm\delta}) & \ddot\psi_{{\bm\beta}{\bm\omega},{\bm\alpha}{\bm\delta}}({\bm\alpha},{\bm\delta}) \cr \ddot\psi_{{\bm\beta}{\bm\omega},{\bm\delta}{\bm\alpha}}({\bm\alpha},{\bm\delta}) & \ddot\psi_{{\bm\beta}{\bm\omega},{\bm\delta}{\bm\delta}}({\bm\alpha},{\bm\delta}) \end{array} \right),
\end{equation}
where $\ddot\psi_{{\bm\beta}{\bm\omega},{\bm\delta}{\bm\alpha}}({\bm\alpha},{\bm\delta})=\ddot\psi_{{\bm\beta}{\bm\omega},{\bm\alpha}{\bm\delta}}^\top({\bm\alpha},{\bm\delta})$. The Hessian matrix $\ddot\psi_{{\bm\beta}{\bm\omega},{\bm\alpha}{\bm\alpha}}({\bm\alpha},{\bm\delta})$ of $\psi_{{\bm\beta}{\bm\omega}}({\bm\alpha},{\bm\delta})$ with respect to ${\bm\alpha}$ is
\begin{equation}
\label{eq:hessian matrix of modified working objective function for MLE for computing for alpha}
\eqalign{
\ddot\psi_{{\bm\beta}{\bm\omega},{\bm\alpha}{\bm\alpha}}({\bm\alpha},{\bm\delta})=&-{\bf X}^\top{\bf R}_{{\bm\beta}{\bm\omega},{\bm\delta}}^{-1}{\bf X}.
}\end{equation}
The $j$th column of the Hessian matrix $\ddot\psi_{{\bm\beta}{\bm\omega},{\bm\alpha}{\bm\omega}}({\bm\alpha},{\bm\delta})$ of $\psi_{{\bm\beta}{\bm\omega}}({\bm\alpha},{\bm\delta})$ with respect to ${\bm\alpha}$ and ${\bm\delta}$ is
\begin{equation}
\label{eq:hessian matrix of modified working objective function for MLE for computing for alpha and delta}
\eqalign{
{\partial \dot\psi_{{\bm\beta}{\bm\omega},{\bm\alpha}}({\bm\alpha},{\bm\delta})\over\partial\delta_j}=&-{\bf X}^\top{\bf R}_{{\bm\beta}{\bm\omega},{\bm\delta}}^{-1}{\partial {\bf R}_{{\bm\beta}{\bm\omega},{\bm\delta}}\over\partial\delta_j}{\bf R}_{{\bm\beta}{\bm\omega},{\bm\delta}}^{-1}(\tilde{\bm y}_{{\bm\beta}\hat{\bm\eta}_{{\bm\beta}{\bm\omega}}}-{\bf X}{\bm\beta}-{\bf X}{\bm\alpha})
}\end{equation}
for $j=1,\dots,r$. The $(j_1,j_2)$th entry of the Hessian matrix $\ddot\psi_{{\bm\beta}{\bm\omega},{\bm\omega}{\bm\omega}}({\bm\alpha},{\bm\delta})$ of $\psi_{{\bm\beta}{\bm\omega}}({\bm\alpha},{\bm\delta})$ with respect to ${\bm\delta}$ is
\begin{equation}
\label{eq:hessian of modified working objective function for MLE for computing for delta}
\eqalign{
{\partial^2\psi_{{\bm\beta}{\bm\omega}}({\bm\alpha},{\bm\delta})\over\partial\delta_{j_1}\partial\delta_{j_2}}=&-{1\over 2}{\rm tr}\left( {\bf R}_{{\bm\beta}{\bm\omega},{\bm\delta}}^{-1}{\partial^2 {\bf R}_{{\bm\beta}{\bm\omega},{\bm\delta}}\over\partial\delta_{j_1}\partial\delta_{j_2}}\right)+{1\over 2}{\rm tr}\left( {\bf R}_{{\bm\beta}{\bm\omega},{\bm\delta}}^{-1}{\partial {\bf R}_{{\bm\beta}{\bm\omega},{\bm\delta}}\over\partial\delta_{j_1}} {\bf R}_{{\bm\beta}{\bm\omega},{\bm\delta}}^{-1}{\partial {\bf R}_{{\bm\beta}{\bm\omega},{\bm\delta}}\over\partial\delta_{j_2}}\right)\cr
& +{1\over 2}(\tilde{\bm y}_{{\bm\beta}\hat{\bm\eta}_{{\bm\beta}{\bm\omega}}}-{\bf X}{\bm\beta}-{\bf X}{\bm\alpha})^\top {\bf R}_{{\bm\beta}{\bm\omega},{\bm\delta}}^{-1}{\partial^2{\bf R}_{{\bm\beta}{\bm\omega},{\bm\delta}}\over\partial\delta_{j_1}\partial\delta_{j_2}}{\bf R}_{{\bm\beta}{\bm\omega},{\bm\delta}}^{-1} (\tilde{\bm y}_{{\bm\beta}\hat{\bm\eta}_{{\bm\beta}{\bm\omega}}}-{\bf X}{\bm\beta}-{\bf X}{\bm\alpha}) \cr
& -(\tilde{\bm y}_{{\bm\beta}\hat{\bm\eta}_{{\bm\beta}{\bm\omega}}}-{\bf X}{\bm\beta}-{\bf X}{\bm\alpha})^\top {\bf R}_{{\bm\beta}{\bm\omega},{\bm\delta}}^{-1}{\partial {\bf R}_{{\bm\beta}{\bm\omega},{\bm\delta}}\over\partial\delta_{j_1}}\cr
&\hspace{2cm}{\bf R}_{{\bm\beta}{\bm\omega},{\bm\delta}}^{-1}{\partial {\bf R}_{{\bm\beta}{\bm\omega},{\bm\delta}}\over\partial\delta_{j_2}}{\bf R}_{{\bm\beta}{\bm\omega},{\bm\delta}}^{-1} (\tilde{\bm y}_{{\bm\beta}\hat{\bm\eta}_{{\bm\beta}{\bm\omega}}}-{\bf X}{\bm\beta}-{\bf X}{\bm\alpha})\cr
}\end{equation}
for $j_1,j_2=1,\dots,r$. 

We use Theorem~\ref{thm:gradient property of modified working objective function for MLE} to design an algorithm for $(\check{\bm\beta}^\top,\check{\bm\omega}^\top)^\top$ defined by~\eqref{eq:MLE for score equation likelihood}. The algorithm does not contain any numerical evaluations of intractable integrals because $\psi_{{\bm\beta}{\bm\omega}}({\bm\alpha},{\bm\delta})$ is a logarithm of a normal density. We treat ${\bm\alpha}$ and ${\bm\delta}$ as variables for the computational goal. We apply the Newton-Raphson approach via $\dot\ell_{{\bm\beta}{\bm\omega}}({\bm 0},{\bm 0})$ and $\ddot\ell_{{\bm\beta}{\bm\omega}}({\bm 0},{\bm 0})$. The computation induces a local optimizer of $\psi_{{\bm\beta}{\bm\omega}}({\bm\alpha},{\bm\delta})$ as
\begin{equation}
\label{eq:newton raphson method for alpha and delta for the next}
\left(\begin{array}{c}{\bm\alpha}_{{\bm\beta}{\bm\omega}}\cr {\bm\delta}_{{\bm\beta}{\bm\omega}}\cr \end{array}\right)=\ddot\psi_{{\bm\beta}{\bm\omega}}^{-1}({\bm 0},{\bm 0})\dot\psi_{{\bm\beta}{\bm\omega}}^{-1}({\bm 0},{\bm 0})
\end{equation}
in the iterations. It then updates $({\bm\beta}^\top,{\bm\omega}^\top)^\top$ by
\begin{equation}
\label{eq:update of beta and omega for the MLE}
\left(\begin{array}{c}{\bm\beta}\cr {\bm\omega}\cr \end{array}\right) \leftarrow \left(\begin{array}{c}{\bm\beta}\cr {\bm\omega}\cr \end{array}\right)+\left(\begin{array}{c}{\bm\alpha}_{{\bm\beta}{\bm\omega}}\cr {\bm\delta}_{{\bm\beta}{\bm\omega}}\cr \end{array}\right)
\end{equation}
for the next iteration. To start the algorithm, we initialize ${\bm\eta}$ with an algorithm for $\hat{\bm\gamma}_{{\bm\beta}{\bm\omega}}$ based on given $({\bm\beta}^\top,{\bm\omega}^\top)^\top$ with the method proposed by \cite{zhang2024}. We obtain Algorithm~\ref{alg:newton-raphson algorithm for the MLE}. 

\begin{algorithm}
\caption{\label{alg:newton-raphson algorithm for the MLE} A solution to $\dot\ell({\bm\beta},{\bm\omega})={\bm 0}$}
\begin{algorithmic}[1]
\Statex{{\bf Input}: Data from the GLMM defined by~\eqref{eq:exponential family distribution},~\eqref{eq:generalized linear mixed model}, and~\eqref{eq:distribution of random effects}}
\Statex{{\bf Output}: $(\check{\bm\beta}^\top,\check{\bm\omega}^\top)^\top$, $\hat{\bm\gamma}_{\check{\bm\beta}\check{\bm\omega}}$, $\tilde{\bm y}_{\check{\bm\beta}\hat{\bm\gamma}_{\check{\bm\beta}\check{\bm\omega}}}$, ${\bf W}_{\check{\bm\beta}\hat{\bm\gamma}_{\check{\bm\beta}\check{\bm\omega}}}$, ${\bf D}_{\check{\bm\omega}}$, ${\bf R}_{\check{\bm\beta}\check{\bm\omega},{\bm 0}}$, $\psi_{\check{\bm\beta}\check{\bm\omega}}({\bm 0},{\bm 0})$, $\dot\psi_{\check{\bm\beta}\check{\bm\omega}}({\bm 0},{\bm 0})$, and $\ddot\psi_{\check{\bm\beta}\check{\bm\omega}}({\bm 0},{\bm 0})$}
\Statex{\it Initialization}
\State{Initialize ${\bm\eta}^{(0)}$ and set $\tilde{\bm y}^{(0)}\leftarrow{\bm\eta}^{(0)}+[{\bm y}-b'({\bm\eta}^{(0)})]/b''({\bm\eta}^{(0)})$ and ${\bf W}^{(0)}\leftarrow{\rm diag}[b''({\bm\eta}^{(0)})]$}
\State{Initialize~\eqref{eq:modified working objective function for MLE for computing} by $\psi^{(0)}({\bm\alpha},{\bm\delta})=\log[\varphi(\tilde{\bm y}^{(0)};{\bf X}{\bm\alpha},{\bf R}_{\bm\delta}^{(0)})]$ with ${\bf R}_{\bm\delta}^{(0)}=\{{\bf W}^{(0)}\}^{-1}+{\bf Z}{\bf D}_{\bm\delta}{\bf Z}^\top$}
\State{${\bm\beta}^{(0)}\leftarrow{\bm\alpha}^{(0)}$ and ${\bm\omega}^{(0)}\leftarrow{\bm\delta}^{(0)}$ by solving $\dot\psi^{(0)}({\bm\alpha}^{(0)},{\bm\delta}^{(0)})={\bm 0}$ using~\eqref{eq:newton raphson method for alpha and delta for the next}}
\State{Compute $\hat{\bm\gamma}^{(0)}=\hat{\bm\gamma}_{{\bm\beta}^{(0)}{\bm\omega}^{(0)}}$ by initializing $\tilde{\bm\eta}={\bm\eta}^{(0)}$, $\tilde{\bm y}=\tilde{\bm y}^{(0)}$, and ${\bf W}={\bf W}^{(0)}$, and then iteratively update ${\bf R}\leftarrow {\bf W}^{-1}+{\bf Z}{\bf D}_{{\bm\omega}^{(0)}}{\bf Z}^\top$, $\hat{\bm\gamma}^{(0)}\leftarrow {\bf D}_{{\bm\omega}^{(0)}}{\bf Z}^\top{\bf R}^{-1}(\tilde{\bm y}-{\bf X}{\bm\beta}^{(0)})$, $\tilde{\bm\eta}={\bf X}{\bm\beta}^{(0)}+{\bf Z}\hat{\bm\gamma}^{(0)}$, $\tilde{\bm y}\leftarrow\tilde{\bm\eta}+[{\bm y}-b'(\tilde{\bm\eta})]/b''(\tilde{\bm\eta})$, and ${\bf W}\leftarrow{\rm diag}[b''(\tilde{\bm\eta})]$}
\Statex{\it Begin Iteration}
\State{${\bm\eta}^{(t)}\leftarrow {\bf X}{\bm\beta}^{(t-1)}+{\bf Z}\hat{\bm\gamma}^{(t-1)}$, $\tilde{\bm y}^{(t)}\leftarrow{\bm\eta}^{(t)}+[{\bm y}-b'({\bm\eta}^{(t)})]/b''({\bm\eta}^{(t)})$, ${\bf W}^{(t)}\leftarrow{\rm diag}[b''({\bm\eta}^{(t)})]$ }
\State{Let $\psi^{(t)}({\bm\alpha},{\bm\delta})=\log[\varphi(\tilde{\bm y}^{(t)};{\bf X}{\bm\beta}^{(t-1)}+{\bf X}{\bm\alpha},{\bf R}_{{\bm\delta}}^{(t)})]$ with ${\bf R}_{{\bm\delta}}^{(t)}=\{{\bf W}^{(t)}\}^{-1}+{\bf Z}{\bf D}_{{\bm\omega}^{(t-1)}+{\bm\delta}}{\bf Z}^\top$}
\State{ ${\bm\beta}^{(t)}\leftarrow{\bm\beta}^{(t-1)}+{\bm\alpha}^{(t)}$ and ${\bm\omega}^{(t)}\leftarrow{\bm\omega}^{(t-1)}+{\bm\delta}^{(t)}$ by solving $\dot\psi^{(t)}({\bm\alpha}^{(t)},{\bm\delta}^{(t)})={\bm 0}$ using~\eqref{eq:newton raphson method for alpha and delta for the next}}
\State{Compute $\hat{\bm\gamma}^{(t)}=\hat{\bm\gamma}_{{\bm\alpha}^{(t)}{\bm\delta}^{(t)}}$ by initializing $\tilde{\bm\eta}={\bm\eta}^{(t)}$, $\tilde{\bm y}=\tilde{\bm y}^{(t)}$, and ${\bf W}={\bf W}^{(t)}$, and then iteratively update ${\bf R}\leftarrow {\bf W}^{-1}+{\bf Z}{\bf D}_{{\bm\omega}^{(t)}}{\bf Z}^\top$, $\hat{\bm\gamma}^{(t)}\leftarrow {\bf D}_{{\bm\omega}^{(t)}}{\bf Z}^\top{\bf R}^{-1}(\tilde{\bm y}-{\bf X}{\bm\beta}^{(t)})$, $\tilde{\bm\eta}={\bf X}{\bm\beta}^{(t)}+{\bf Z}\hat{\bm\gamma}^{(t)}$, $\tilde{\bm y}\leftarrow\tilde{\bm\eta}+[{\bm y}-b'(\tilde{\bm\eta})]/b''(\tilde{\bm\eta})$, and ${\bf W}\leftarrow{\rm diag}[b''(\tilde{\bm\eta})]$}
\Statex{\it End Iteration}
\State {Output}
\end{algorithmic}
\end{algorithm}

\begin{cor}
\label{cor:solution of local} 
The final answer of $({\bm\beta}^{(t)\top},{\bm\omega}^{(t)\top})^\top$ given by Algorithm~\ref{alg:newton-raphson algorithm for the MLE} belongs to ${\mathcal A}$. Any $(\check{\bm\beta}^\top,\check{\bm\omega}^\top)^\top\in{\mathcal A}$ can be a final answer of Algorithm~\ref{alg:newton-raphson algorithm for the MLE}.
\end{cor}

Corollary~\ref{cor:solution of local} points out that~\eqref{eq:MLE for score equation likelihood} can be exactly solved by Algorithm~\ref{alg:newton-raphson algorithm for the MLE}. The computation does not involve any numerical evaluations of intractable integrals contained on the right-hand side of~\eqref{eq:likelihood function marginal}. The final answers can only be treated as a local optimizer of~\eqref{eq:MLE of parameters and hyperparameters}. This is caused by the usage of~\eqref{eq:newton raphson method for alpha and delta for the next} in Steps 3 and 6 because $({\bm\alpha}_{{\bm\beta}{\bm\delta}}^\top,{\bm\delta}_{{\bm\beta}{\bm\delta}}^\top)^\top$ is a local optimizer of $\psi_{{\bm\beta}{\bm\omega}}({\bm\alpha},{\bm\delta})$. The difficulty can be overcome by a modification of Steps 3 and 6 of Algorithm~\ref{alg:newton-raphson algorithm for the MLE} as
\begin{equation}
\label{eq:global optimizer for working loglikelihood}
(\hat{\bm\alpha}_{{\bm\beta}{\bm\omega}}^\top,\hat{\bm\delta}_{{\bm\beta}{\bm\omega}}^\top)^\top=\mathop{\arg\!\max}_{{\bm\alpha},{\bm\delta}}\psi_{{\bm\beta}{\bm\omega}}({\bm\alpha},{\bm\delta})
\end{equation}
with a modification of~\eqref{eq:update of beta and omega for the MLE} correspondingly. We compare~\eqref{eq:global optimizer for working loglikelihood} with the second-order LA given in~\eqref{eq:laplace approximation formula}. We find that $\psi_{{\bm\beta}{\bm\omega}}({\bm\alpha},{\bm\delta})$ can be treated as the logarithm of the second-order approximation of $L({\bm\beta},{\bm\omega}|{\bm\gamma})$ at ${\bm\gamma}=\hat{\bm\gamma}_{\check{\bm\beta}\check{\bm\omega}}$ for some $(\check{\bm\beta},\check{\bm\omega}^\top)^\top\in{\mathcal A}$. Using this property, we compare the values of $\psi_{\check{\bm\beta}\check{\bm\omega}}({\bm 0},{\bm 0})$ for all $(\check{\bm\beta},\check{\bm\omega}^\top)^\top\in{\mathcal A}$ in deriving a solution to~\eqref{eq:MLE of parameters and hyperparameters}. The approach has been previously used in the LA for computing an approximate MLE of a GLMM. Besides computational advantages, our method provides the exact MLE but the LA method only provides an approximate MLE. We specify our method to the following two typical GLMMs.

{\it Example 1.} The binomial GLMM under the logistic link is formulated as ${\bm y}_i|{\bm\gamma}\sim{\mathcal Bin}(m_i,\pi_i)$ independently with $\log[\pi_i/(1-\pi_i)]={\bm x}_i^\top{\bm\beta}+{\bm z}_i{\bm\gamma}$ for $i=1,\dots,n$. We initialize $\eta_i^{(0)}=\log[(y_i+0.5)/(m_i-y_i+0.5)]$ in Step 1. By $b(\eta_i^{(0)})=m_i\log[1+\exp(\eta_i^{(0)})]$, we obtain $b'(\eta_i^{(0)})=m_i\exp(\eta_i^{(0)})/[1+\exp(\eta_i^{(0)})]$ and $b''(\eta_i^{(0)})=m_i\exp(\eta_i^{(0)})/[1+\exp(\eta_i^{(0)})]^2$, leading to $\tilde y_i^{(0)}=\log[(y_i+0.5)/(m_i-y_i+0.5)]+(y_i-0.5m_i)(m_i+1)/[m_i(y_i+0.5)(m_i-y_i+0.5)]$ and $w_i^{(0)}=m_i(y_i+0.5)(m_i-y_i+0.5)/(m_i+1)^2$ also in Step 1 of Algorithm~\ref{alg:newton-raphson algorithm for the MLE}. The objective function in Step 2 given by
\begin{equation}
\label{eq:objective function initial binomial}
\psi^{(0)}({\bm\alpha},{\bm\delta})=-{1\over 2}\log(2\pi)-{1\over 2}\log|\det({\bf R}_{\bm\delta}^{(0)})|-{1\over 2}(\tilde{\bm y}^{(0)}-{\bf X}{\bm\alpha})^\top\{{\bf R}_{\bm\delta}^{(0)}\}^{-1} (\tilde{\bm y}^{(0)}-{\bf X}{\bm\alpha})
\end{equation}
with ${\bf R}_{\bm\delta}^{(0)}={\bf W}^{(0)}+{\bf Z}^\top{\bf D}_{\bm\delta}{\bf Z}$ and ${\bf W}^{(0)}={\rm diag}(w_1^{(0)},\dots,w_n^{(0)})$ can be applied. A solution ${\bm\beta}^{(0)}$ and ${\bm\omega}^{(0)}$ in Step 3 is derived by solving $\dot\psi^{(0)}({\bm\beta}^{(0)},{\bm\omega}^{(0)})={\bm 0}$ numerically. With ${\bm\beta}^{(0)}$ and ${\bm\omega}^{(0)}$, the computation of $\hat{\bm\gamma}^{(0)}$ in Step 4 can be straightforwardly applied. Step 5 assumes that $\hat{\bm\gamma}^{(t-1)}$, ${\bm\beta}^{(t-1)}$, and ${\bm\omega}^{(t-1)}$ have been derived in the previous iteration. It starts with ${\bm\eta}^{(t)}={\bf X}{\bm\beta}^{(t-1)}+{\bf Z}{\bm\gamma}^{(t-1)}$ and computes $\tilde{\bm y}^{(t)}={\bm\eta}^{(t)}+({\bm y}-{\bm m}\circ {\bm\pi}^{(t)})/[{\bm m}\circ{\bm\pi}^{(t)}\circ({\bm 1}-{\bm\pi}^{(t)})]$ and ${\bf W}^{(t)}={\bm m}\circ{\bm\pi}^{(t)}\circ({\bm 1}-{\bm\pi}^{(t)})$, where ${\bm\pi}^{(t)}=\exp({\bm\eta}^{(t)})/[{\bm 1}+\exp({\bm\eta}^{(t)})]$, ${\bm m}=(m_1,\dots,m_n)^\top$, and $\circ$ represents the Hadamard (i.e., elementwise) product of vectors or matrices. The objective function in Step 6 as
\begin{equation}
\label{eq:objective function iteration binomial}
\psi^{(t)}({\bm\alpha},{\bm\delta})=-{1\over 2}\log(2\pi)-{1\over 2}\log|\det({\bf R}_{\bm\delta}^{(t)})|-{1\over 2}(\tilde{\bm y}^{(t)}-{\bf X}{\bm\beta}-{\bf X}{\bm\alpha})^\top\{{\bf R}_{\bm\delta}^{(t)}\}^{-1} (\tilde{\bm y}^{(t)}-{\bf X}{\bm\beta}-{\bf X}{\bm\alpha})
\end{equation}
with ${\bf R}_{\bm\delta}^{(t)}={\bf W}^{(t)}+{\bf Z}^\top{\bf D}_{{\bm\omega}^{(t-1)}+{\bm\delta}}{\bf Z}$ can be calculated by matrix operations. The solution ${\bm\alpha}^{(t)}$ and ${\bm\delta}^{(t)}$ in Step 7 is derived by solving $\dot\psi^{(t)}({\bm\alpha}^{(t)},{\bm\delta}^{(t)})={\bm 0}$ numerically. It updates the estimates of ${\bm\beta}$ and ${\bm\omega}$ with ${\bm\beta}^{(t)}={\bm\beta}^{(t-1)}+{\bm\alpha}^{(t)}$ and ${\bm\omega}^{(t)}={\bm\omega}^{(t-1)}+{\bm\delta}^{(t)}$, respectively. Using ${\bm\beta}^{(t)}$ and ${\bm\omega}^{(t)}$, the computation of $\hat{\bm\gamma}^{(t)}$ in Step 8 can be applied. It supplies $\hat{\bm\gamma}^{(t)}$, ${\bm\beta}^{(t)}$, and ${\bm\omega}^{(t)}$ for the next iteration. All steps of Algorithm~\ref{alg:newton-raphson algorithm for the MLE} are derived. The exact values of $\check{\bm\beta}$ and $\check{\bm\omega}$ are contained in the final answers. The global solution to~\eqref{eq:MLE of parameters and hyperparameters} can be found by searching the maximum of the final values of $\psi^{(t)}({\bm 0},{\bm 0})$ for all possible $(\check{\bm\beta},\check{\bm\omega})^\top$ given by the algorithm.  \qed

{\it Example 2.} The Poisson GLMM under the log link is formulated as $y_i|{\bm\gamma}\sim{\mathcal P}(\mu_i)$ independently with $\log(\mu_i)={\bm x}_i^\top{\bm\beta}+{\bm z}_i{\bm\gamma}$ for $i=1,\dots,n$. We initialize $\eta_i^{(0)}=\log(y_i+0.5)$. By $b(\eta_i^{(0)})=\exp(\eta_i^{(0)})$, we obtain $b'(\eta_i^{(0)})=b''(\eta_i^{(0)})=\exp(\eta_i^{(0)})$, leading to $\tilde y_i^{(0)}=\log(y_i+0.5)+0.5/(y_i+0.5) $ and $w_i^{(0)}=y_i+0.5$ in Step 1 of Algorithm~\ref{alg:newton-raphson algorithm for the MLE}. The objective function in Step 2 is calculated identically by~\eqref{eq:objective function initial binomial}, leading to a solution ${\bm\beta}^{(0)}$ and ${\bm\omega}^{(0)}$ in Step 3 derived similarly. The computation of $\hat{\bm\gamma}^{(0)}$ in Step 4 is also similar. Given that $\hat{\bm\gamma}^{(t-1)}$, ${\bm\beta}^{(t-1)}$, and ${\bm\omega}^{(t-1)}$ have been derived in the previous iteration, Step 5 is carried out by setting ${\bm\eta}^{(t)}={\bf X}{\bm\beta}^{(t-1)}+{\bf Z}{\bm\gamma}^{(t-1)}$, $\tilde{\bm y}^{(0)}={\bm\eta}^{(t)}+[{\bm y}-\exp({\bm\eta}^{(t)})]/\exp({\bm\eta}^{(t)})$, and ${\bf W}^{(t)}={\rm diag}[\exp({\bm\eta}^{(0)})]$. The objective function in Step 6 is computed identically by~\eqref{eq:objective function iteration binomial}, leading to a solution of ${\bm\beta}^{(t)}$ and ${\bm\omega}^{(t)}$ in Step 7, and $\hat{\bm\gamma}^{(t)}$ in Step 8. The exact values of $\check{\bm\beta}$ and $\check{\bm\omega}$ are contained in the final answers of the algorithm. A solution to~\eqref{eq:MLE of parameters and hyperparameters} can be similarly derived.  \qed

Examples 1 and 2 indicate that we can exactly solve~\eqref{eq:MLE of parameters and hyperparameters} and~\eqref{eq:MLE for score equation likelihood} in the binomial and Poisson GLMMs. The previous methods developed under the LA or Monte Carlo approaches can only provide approximate solutions. The objective functions $\psi^{(0)}({\bm\alpha},{\bm\delta})$ and $\psi^{(t)}({\bm\alpha},{\bm\delta})$ in Steps 2 and 6 of Algorithm~\ref{alg:newton-raphson algorithm for the MLE} are computed by matrix operations. The predictions of the random effects given by $\hat{\bm\gamma}^{(0)}$ and $\hat{\bm\gamma}^{(t)}$ in Steps 4 and 8 are also carried out by matrix operations. Therefore, Algorithm~\ref{alg:newton-raphson algorithm for the MLE} can be treated as a PM algorithm proposed by~\cite{zhang2023}. It has prediction and maximization steps. Computation in the PM algorithm does not involve any numerical evaluations of intractable integrals for the right-hand side of~\eqref{eq:likelihood function marginal}. We avoid the main difficulty in computing the exact MLE of $({\bm\beta}^\top,{\bm\omega}^\top)^\top$ for the cases when ${\bm y}|{\bm\gamma}$ does not follow normal. We solve the exact MLE problem defined by the right-hand sides of~\eqref{eq:MLE of parameters and hyperparameters} and~\eqref{eq:MLE for score equation likelihood}. The exact likelihood function given by~\eqref{eq:likelihood function marginal} remains unsolved. 

\section{Test Statistics}
\label{sec:test statistics}

We formulate the likelihood ratio, the score, and the generalized Wald statistics based on the output of Algorithm~\ref{alg:newton-raphson algorithm for the MLE}. The goal is to measure the difference between two estimates of $({\bm\beta}^\top,{\bm\omega}^\top)^\top$, denoted as $(\hat{\bm\beta}_1^\top,\hat{\bm\omega}_1^\top)^\top$ and $(\hat{\bm\beta}_2^\top,\hat{\bm\omega}_2^\top)^\top$ for statistical models ${\mathcal M}_1$ and ${\mathcal M}_2$, respectively. We assume that ${\mathcal M}_1$ is a special case of ${\mathcal M}_2$. We test whether ${\mathcal M}_2$ can be reduced to ${\mathcal M}_1$ under the null hypothesis ($H_0$) as
\begin{equation}
\label{eq:null and alternative hypotheses}
\left(\begin{array}{c}{\bm\beta}_1\cr {\bm\omega}_1\cr \end{array} \right)={\bf B}\left(\begin{array}{c}{\bm\beta}_2\cr {\bm\omega}_2 \cr\end{array} \right),
\end{equation}
where ${\bf B}\in\mathbb{R}^{k_1\times k_2}$ with $k_1<k_2$ and ${\rm rank}({\bf B})=k_1$ is the matrix for the relationship between ${\mathcal M}_1$ and ${\mathcal M}_2$. Given the left-hand side of~\eqref{eq:null and alternative hypotheses}, the right-hand side can be solved by 
\begin{equation}
\label{eq:special case in full model based on the reduced model}
\left(\begin{array}{c}{\bm\beta}_2^*\cr {\bm\omega}_2^* \cr\end{array} \right)={\bf B}^* \left(\begin{array}{c}{\bm\beta}_1\cr {\bm\omega}_1\cr \end{array} \right),
\end{equation}
where ${\bf B}^*={\bf V}{\bf D}^{-1}{\bf U}^\top$ is the Moore–Penrose generalized inverse derived based on the singular value decomposition (SVD) of ${\bf B}$ as ${\bf B}={\bf U}{\bf D}{\bf V}^\top$~\cite{zhang2016}. The solution given by the left-hand side of~\eqref{eq:special case in full model based on the reduced model} is unique.

The likelihood ratio statistic is defined as
\begin{equation}
\label{eq:likelihood ratio statistics}
\Lambda= 2[\ell(\hat{\bm\beta}_2,\hat{\bm\omega}_2)-\ell(\hat{\bm\beta}_1,\hat{\bm\omega}_2)]=2[\log\bar f_{\hat{\bm\beta}_2\hat{\bm\omega}_1}({\bm y})-\log\bar f_{\hat{\bm\beta}_1\hat{\bm\omega}_1}({\bm y})].
\end{equation}
There is $\Lambda\ge 0$ in~\eqref{eq:likelihood ratio statistics}. Because the exact likelihood function given by~\eqref{eq:likelihood function marginal} remains unsolved, we cannot compute the exact value of $\Lambda$. We approximate the likelihood ratio statistic as
\begin{equation}
\label{eq:approximate likelihod ratio statistic}
\Lambda_{\psi}=2[\psi_{\hat{\bm\beta}_2\hat{\bm\omega}_2}({\bm 0},{\bm 0})-\psi_{\hat{\bm\beta}_1\hat{\bm\omega}_1}({\bm 0},{\bm 0})].
\end{equation}
It is devised based on the property that the objective function $\psi_{\check{\bm\beta}\check{\bm\omega}}({\bm 0},{\bm 0})$ can be treated as an approximation of the loglikelihood function at the exact solution $(\check{\bm\beta}^\top,\check{\bm\omega}^\top)^\top$ to~\eqref{eq:MLE for score equation likelihood}. The approach is common in the LA approach for approximating the likelihood function. 

We need the Fisher Information in constructing the score and the generalized Wald statistics. The Fisher Information at a solution $(\check{\bm\beta}^\top,\check{\bm\omega}^\top)^\top$ to~\eqref{eq:MLE for score equation likelihood} is defined as 
\begin{equation}
\label{eq:fisher information}
\eqalign{
{\mathcal I}({\check{\bm\beta},\check{\bm\omega}})=&-{\rm E}_{\check{\bm\beta}\check{\bm\omega}}[\ddot\ell(\check{\bm\beta},\check{\bm\omega})]={\rm E}_{\check{\bm\beta}\check{\bm\omega}}[\dot\ell(\check{\bm\beta},\check{\bm\omega})\dot\ell^\top(\check{\bm\beta},\check{\bm\omega})]\cr
=& \int_{\mathbb{R}^n}  [\dot\ell(\check{\bm\beta},\check{\bm\omega})\dot\ell^\top(\check{\bm\beta},\check{\bm\omega})] \bar f_{\check{\bm\beta}\check{\bm\omega}}({\bm y})\nu(d{\bm y}),
}
\end{equation}
where $\nu(d{\bm y})$ is the counting measure if ${\bm y}$ is discrete or the Lebesgue measure if ${\bm y}$ is continuous. Because $\bar f_{\check{\bm\beta}\check{\bm\omega}}({\bm y})$ given by~\eqref{eq:likelihood function marginal} remains unsolved, we cannot use~\eqref{eq:fisher information} to compute the Fisher Information. Note that $-\ddot\ell(\check{\bm\beta},\check{\bm\omega})$ is often treated as a sample Fisher Information in practice. By $\dot\psi_{\check{\bm\beta}\check{\bm\omega}}({\bm 0},{\bm 0})=\dot\ell(\check{\bm\beta},\check{\bm\omega})$, there is ${\mathcal I}({\check{\bm\beta},\check{\bm\omega}})={\rm E}_{\check{\bm\beta}\check{\bm\omega}}[\dot\psi_{\check{\bm\beta}\check{\bm\omega}}({\bm 0},{\bm 0})\dot\psi_{\check{\bm\beta}\check{\bm\omega}}^\top({\bm 0},{\bm 0})]$, implying that we can treat
\begin{equation}
\label{eq:sample Fisher Information}
{\mathcal I}_{\psi}({\check{\bm\beta},\check{\bm\omega}})=-\ddot\psi_{\check{\bm\beta}\check{\bm\omega}}({\bm 0},{\bm 0})
\end{equation}
as a sample Fisher Information. An advantage is that the right-hand side of~\eqref{eq:sample Fisher Information} is contained in the output of Algorithm~\ref{alg:newton-raphson algorithm for the MLE}. It can be easily implemented. 

Let $(\hat{\bm\beta}_2^{*\top},\hat{\bm\omega}_2^{*\top})^\top$ be the solution given by the left-hand side of~\eqref{eq:special case in full model based on the reduced model} under $(\hat{\bm\beta}_1^\top,\hat{\bm\omega}_1^\top)^\top$ on the right-hand side. Based on~\eqref{eq:sample Fisher Information}, we devise our score statistic as
\begin{equation}
\label{eq:score statistic}
S_{\psi}= -\dot\psi_{\hat{\bm\beta}_2^*\hat{\bm\omega}_2^*}^\top({\bm 0},{\bm 0})\ddot\psi_{\hat{\bm\beta}_2^*\hat{\bm\omega}_2^*}^{-1}({\bm 0},{\bm 0})\dot\psi_{\hat{\bm\beta}_2^*\hat{\bm\omega}_2^*}({\bm 0},{\bm 0})
\end{equation}
and our generalized Wald statistic as
\begin{equation}
\label{eq:generalized wald statistic}
GW_{\psi}=  -\left(\begin{array}{cc}\hat{\bm\beta}_2^{*\top} & \hat{\bm\omega}_2^{*\top}\cr  \end{array}\right)           \ddot\psi_{\hat{\bm\beta}_2^*\hat{\bm\omega}_2^*}({\bm 0},{\bm 0}) \left(\begin{array}{c}\hat{\bm\beta}_2^*\cr \hat{\bm\omega}_2^*\cr \end{array}\right).
\end{equation}

Under $H_0$, the likelihood statistic $\Lambda_{\psi}$, the score statistic $S_{\psi}$, and the generalized Wald statistic $GW_{\psi}$ approximately follow the $\chi_{k_2-k_1}^2$ distribution. It is used in computing the $p$-values. In addition to testing the null hypothesis given by~\eqref{eq:null and alternative hypotheses}, they can be used to measure the difference between two different estimation methods. This approach is used in Section~\ref{sec:application}. 

\section{Simulation}
\label{sec:simulation}

We evaluate the performance of our method with the comparison to our competitors via Monte Carlo simulations with $1000$ replications. We specify the GLMM jointly defined by~\eqref{eq:exponential family distribution}, \eqref{eq:generalized linear mixed model}, and~\eqref{eq:distribution of random effects} to a spatial GLMM (SGLMM) for Poisson data as $y_i|{\bm\gamma}\sim{\mathcal P}(\mu_i)$ conditionally independently with 
\begin{equation}
\label{eq:spatial poisson model simulation}
\log\mu_i=\beta_0+x_{i1}\beta_1+x_{i2}\beta_2+\gamma_i
\end{equation}
for $i=1,\dots,400$, implying that we had $n=400$ and $p=3$ in our simulation setting. We generate ${\bm\gamma}=(\gamma_1,\dots,\gamma_{400})^\top$ from~\eqref{eq:distribution of random effects} with the $(i,j)$th entry of ${\bf D}_{\bm\omega}$ specified by the spatial Mat\'ern covariance model as
\begin{equation}
\label{eq:matern family}
M_{\bm\omega}( d_{ij} )={\omega_1\over 1-\omega_1}{(\omega_2d_{ij})^{\omega_3}\over 2^{\omega_3-1}\Gamma(\omega_3)}K_{\omega_3}(\omega_2d_{ij}),
\end{equation}
where ${\bm\omega}=(\omega_1,\omega_2,\omega_3)^\top$ with $\omega_1\in(0,1)$, $\omega_2>0$, and $\omega_3>0$ is the hyperparameter vector, $d_{ij}$ is the distance between the sites $i$ and $j$, $K_{\omega_3}(\cdot)$ is the modified Bessel function of the second kind. The Mat\'ern family is spatially isotropic with the variance, scale, and smoothness hyperparameters as $\omega_1/(1-\omega_1)$, $1/\omega_2$, and $\omega_3$, respectively.  It can be specified to the exponential covariance function by taking $\omega_3=0.5$ in~\eqref{eq:matern family}. The Mat\'ern family was first proposed by~\cite{matern1986} and has received much attention since the theoretical work of~\cite{handcock1993,stein1999}. 

SGLMMs can model purely spatial and spatiotemporal data.  For purely spatial data, $d_{ij}$ is often chosen as the Euclidean distance between sites. For spatiotemporal data, $d_{ij}$ is often represented by a bivariate vector for the spatial and temporal distances. To ensure the variance component matrix is positive definite, parametric families are often used. Examples include the Mat\'ern~\cite{matern1986} and the generalized Cauchy~\cite{gneiting2004} families for purely spatial data. We adopt the well-known Mat\'ern covariance family for purely spatial data in~\eqref{eq:spatial poisson model simulation}.

SGLMMs belong to the hardest category of GLMMs because the dimension of intractable integrals on the right-hand side of~\eqref{eq:likelihood function marginal} increases with $n$. This is different from the GLMMs for longitudinal data, where dependencies are only present for observations within clusters but not between clusters. For an SGLMM, dependencies have to be considered between all observations, leading to difficulty in the LA and the MCMC approaches for the high-dimensional intractable integrals~\cite{best1999,shun1995}. Because the sizes of the clusters are usually small, the LA and the MCMC approaches are reliable in longitudinal data. They may not be reliable in the SGLMMs due to the high-dimensional intractable integrals.

\begin{table}
\caption{\label{tab:RMSE for the comparison of simulation} Simulations with $1000$ replications for the RMSEs of the estimators provided by the proposed method with the comparison to the previous PQL, LA, and INLA methods under the spatial Poisson data generated from the SGLMM defined by~\eqref{eq:spatial poisson model simulation} and~\eqref{eq:matern family}, where the Oracle method represents the ground truth.}
\begin{center}
\begin{tabular}{cccccc}\hline
       &  \multicolumn{5}{c}{RMSE of Estimators}   \\\cline{2-6}
Method &  $\hat\beta_0$ & $\hat\beta_1$ & $\hat\beta_2$ & $\hat\omega_1$ & $\hat\omega_2$ \\\hline
       & \multicolumn{5}{c}{$\beta_1=\beta_2=0.0$}  \\
Oracle &  $0$ & $0$ & $0$ & $0.025$ & $0.140$        \\
Proposed &  $0.109$ & $0.031$ & $0.029$ & $0.025$ & $0.136$      \\
PQL  &  $0.112$ & $0.031$ & $0.029$ & $0.027$ & $0.147$   \\
LA &  $0.112$ & $0.031$ & $0.029$ & $0.028$ & $0.146$    \\
INLA & $5.253$ & $0.052$ & $0.052$ & $0.219$ & $1.000$  \\
       & \multicolumn{5}{c}{$\beta_1=\beta_2=1.0$}  \\
Oracle &  $0$ & $0$ & $0$ & $0.022$ & $0.118$   \\
Proposed & $0.102$ & $0.028$ & $0.033$ & $0.022$ & $0.115$   \\
PQL & $0.367$ & $0.033$ & $0.035$ & $0.414$ & $0.923$ \\
LA & $0.355$ & $0.033$ & $0.035$ & $0.407$ & $0.917$ \\
INLA & $5.282$ & $0.104$ & $0.112$ & $0.223$ & $1.000$ \\
       & \multicolumn{5}{c}{$\beta_1=\beta_2=2.0$}  \\
Oracle &  $0$ & $0$ & $0$ & $0.026$ & $0.128$   \\
Proposed & $0.108$ & $0.028$ & $0.032$ & $0.025$ & $0.128$   \\
PQL &$0.371$ & $0.586$ & $0.574$ & $0.429$ & $31.903$ \\
LA & $0.365$ & $0.591$ & $0.579$ & $0.428$ & $33.250$ \\
INLA & $5.190$ & $0.238$ & $0.231$ & $0.202$ & $1.000$ \\
       & \multicolumn{5}{c}{$\beta_1=\beta_2=3.0$}  \\
Oracle & $0$ & $0$ & $0$ & $0.024$ & $0.129$    \\
Proposed &  $0.111$ & $0.032$ & $0.031$ & $0.024$ & $0.133$ \\
PQL & NA & NA & NA & NA & NA\\
LA & NA & NA & NA & NA & NA\\
INLA & NA & NA & NA & NA & NA\\\hline
\end{tabular}
\end{center}
\end{table}

We generated the spatial Poisson data from~\eqref{eq:spatial poisson model simulation} and~\eqref{eq:matern family} with fixed $\beta_0=10.0$, $\omega_1=0.5$, $\omega_2=1.0$, and $\omega_3=0.5$, and varied $\beta_1=\beta_2=0.0,1.0,2.0,3.0$, leading to $M_{\bm\omega}(d_{ij})=\exp(-d_{ij})$ as the exponential covariance function between the sites. In each replication, we generated $x_{i1}$ and $x_{i2}$ independently from ${\mathcal N}(0,1)$ and locations of sites independently from the uniform distribution on $[0,20]^2$. We computed the fixed effect components by $10.0+\beta_1x_{i1}+\beta_2x_{i2}$ for $i=1,\dots,400$ and the random effects components by ${\bm\gamma}\sim{\mathcal N}({\bm 0},{\bf D}_{\bm\omega})$ with ${\bf D}_{\bm\omega}$ specified by the spatial Mat\'ern covariance model. Based on the linear component at site $i$ as $\eta_i= 10.0+\beta_1x_{i1}+\beta_2x_{i2}+\gamma_i$, we conditionally independently generated the response from $y_i|{\bm\gamma}\sim{\mathcal P}(e^{\eta_i})$ for $i=1,\dots,400$.

We applied our proposed method to the generated data. We compared it with a few previous methods. We considered six previous methods. They were the PQL~\cite{breslow1993}, the LA~\cite{evangelou2011}, the INLA~\cite{rue2009}, the geoRglm~\cite{christensen2002}, the PrevMap~\cite{giorgi2017}, and the geoCount~\cite{jing2015}. In the literature, estimation of parameters and hyperparameters of GLMMs can be classified into likelihood-based or Bayesian estimation. In the category of likelihood-based estimation, the PQL is a restricted maximum likelihood (REML) method. The LA and the PM are the maximum likelihood (ML) methods. The PrevMap is a Monte Carlo maximum likelihood (MCML) method. In the category of Bayesian estimation, the INLA is an approximate Bayesian computation (ABC) method. The geoCount and geoRglm are MCMC methods. We implemented these methods via the corresponding packages of \textsf{R}. The applications of the PrevMap, geoCount, and geoRglm methods failed due to the presence of the regression effects of $x_{i1}$ and $x_{i2}\beta_2$ in the linear components of the SGLMM. The implementation succeeded when the two regression effects were removed from the three methods. We considered the PQL, the LA, and the INLA and discarded the PrevMap, the geoCount, and the geoRglm methods in the comparison. 

In addition, we designed an Oracle method to represent the ground truth of the simulation. The Oracle method assumed that an Oracle observed ${\bm\gamma}$ and used it to estimate ${\bm\omega}$ by maximizing $\varphi({\bm\gamma};{\bm 0},{\bf D}_{\bm\omega})$. It was first proposed for variable selection problems by~\cite{fanli2001} based on the assumption that an Oracle knows the ground truth and can also use it in the computation. The goal is to provide the optimal situation that a statistical method can reach. 

After generating data, we applied the Oracle, our proposed, and the previous PQL, LA, and INLA methods. We assumed that ${\bm\gamma}$ could be used in the Oracle method but not the remaining methods. We also investigated the estimates of ${\bm\beta}$ and ${\bm\omega}$, denoted as $\hat{\bm\beta}=(\hat\beta_0,\hat\beta_1,\hat\beta_2)^\top$ and $\hat{\bm\omega}=(\hat\omega_1,\hat\omega_2,0.5)^\top$, respectively. We treated $\omega_3$ as a constant because we found that the impact of $\omega_3$ could be ignored even if it was treated as a hyperparameter. We then decided to fix $\omega_3=0.5$ in our comparison. 

We used the root mean squared error (RMSE) to measure the accuracy of the estimators. The RMSE values were calculated by the squared root of the sample mean of $\|\beta_0-\hat\beta_0\|^2$, $\|\beta_1-\hat\beta_1\|^2$, $\|\beta_2-\hat\beta_2\|^2$, $\|\omega_1-\hat\omega_1\|^2$, and $\|\omega_2-\hat\omega_2\|^2$, denoted as ${\rm RMSE}(\hat\beta_0)$, ${\rm RMSE}(\hat\beta_1)$, ${\rm RMSE}(\hat\beta_2)$, ${\rm RMSE}(\hat\omega_1)$, and ${\rm RMSE}(\hat\omega_2)$, respectively. The Oracle method did not estimate $\beta_0$, $\beta_1$, and $\beta_2$. It had ${\rm RMSE}(\hat\beta_0)={\rm RMSE}(\hat\beta_1)={\rm RMSE}(\hat\beta_2)=0$, ${\rm RMSE}(\hat\omega_1)>0$, and${\rm RMSE}(\hat\omega_2)>0$. The remaining methods had all positive RMSE values.

We compared our method with the previous PQL, LA, and INLA methods with the Oracle method as the ground truth (Table~\ref{tab:RMSE for the comparison of simulation}). The simulation results showed that the proposed method was the best. It could be as precise as the Oracle method in all the cases studied. The performance of the previous PQL and LA methods was close to our and the Oracle methods when the slopes were absent. The performance became worse when $\beta_1$ and $\beta_2$ increased. The applications of the PQL and the LA failed when $\beta_1=\beta_2=3.0$.  The INLA method was imprecise and the results were unreliable in all the cases studied. 

We also evaluated the computational time.  Our method took around $20$ seconds in computing $\hat{\bm\beta}$ and $\hat{\bm\omega}$ in all the cases studied. The PQL and LA methods took around $20$ seconds when $\beta_1=\beta_2=0.0$. The computation increased to a few minutes when $\beta_1=\beta_2=2.0$. The INLA method took around $25$ seconds in all the cases studied. Because it was imprecise, we do not recommend the INLA method in practice. The simulations suggest using our method. The reason is obvious because our method is the exact MLE but the previous methods can only be treated as the approximate MLE. The approximation provided by the PQL and LA methods is bad if the regression coefficients for explanatory variables is not small. 

\section{Application}
\label{sec:application}

\begin{figure}
\centerline{\rotatebox{270}{\psfig{figure=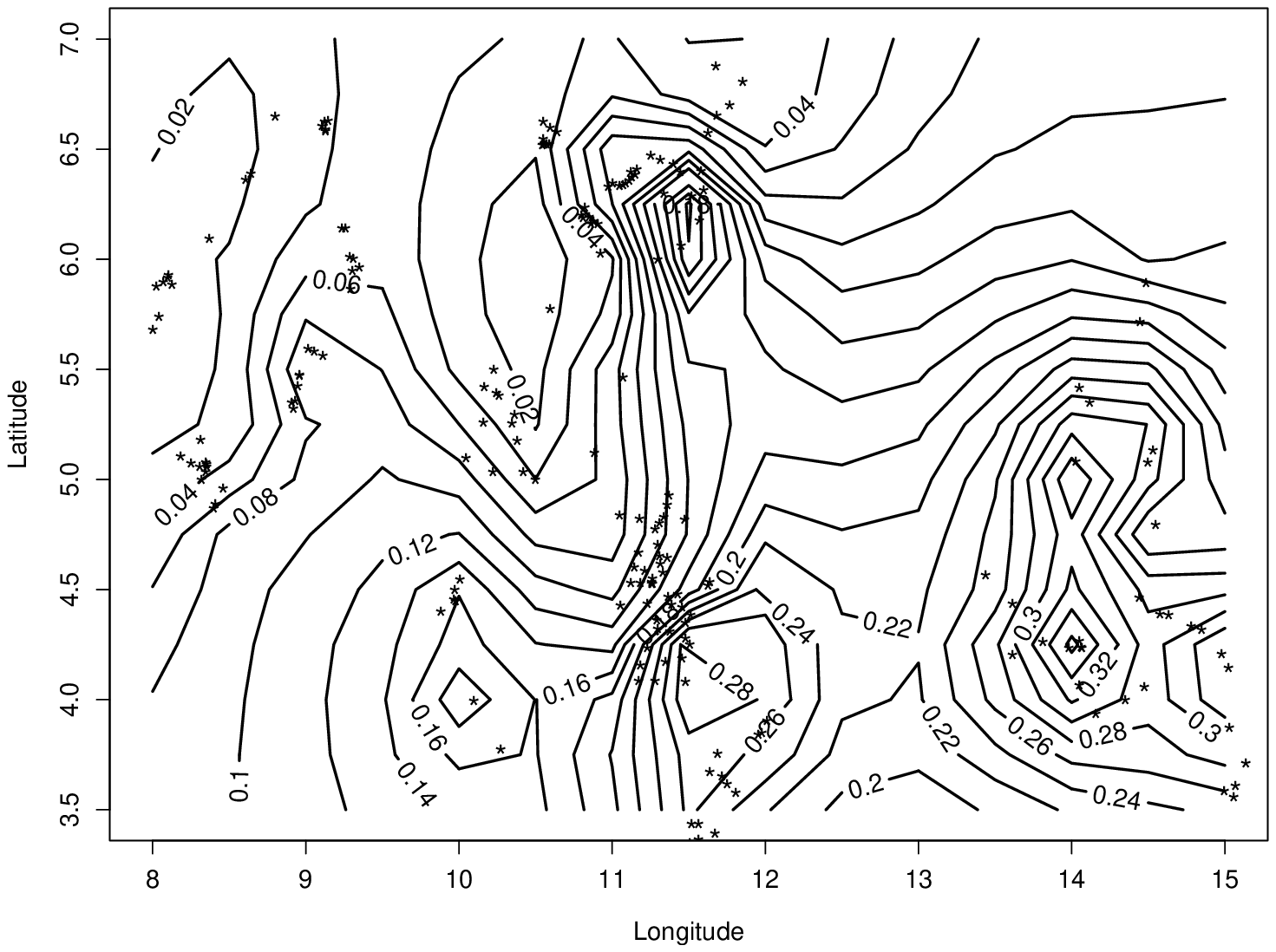,width=3.0in,}}}
\caption{Initial study for the probabilities of \textsf{Loa loa} infections by the Gaussian kriging based on the logistic values of the observed proportions, where the locations of \textsf{Loa loa} samples are marked as *.}
\label{fig:loaloa locations}
\end{figure}

We applied our method to the {\it Loaloa} data. We also considered our competitors. The dataset was previously analyzed by~\cite{diggle2007} and is available in the \textsf{PrevMap} package of \textsf{R}. It represents a study of the prevalence of Loa loa (eyeworm) in a series of surveys undertaken in the $197$ villages in west Africa (Cameroon and southern Nigeria) from 1999 to 2001. In the period of 1999 and 2001, {\it Loa loa} emerged as a filarial worm of significant public-health importance, as a consequence of its impact on the African Programme for Onchocerciasis (APOC).  Severe, sometimes fatal, encephalopathic reactions to ivermectin (the drug of choice for onchocerciasis control) occurred in some individuals with high {\it Loa loa} microfilaraemias.

The {\it Loaloa} dataset has $197$ rows and $11$ columns. The first and the second columns are excluded from the application because they stand for the row ID and the village code, respectively. The locations of the villages are described by their latitudes varied from $3.35^o{\rm N}$ to $6.88^o{\rm N}$ and their longitudes varied from $8.00^o{\rm E}$ to $15.14^o{\rm E}$. The prevalence of Loa loa is described by the {\it number of positive test} (\textsf{NO\_INF} as ${\bm y}=(y_1,\dots,y_{197})^\top$) and the {\it number of people tested} (\textsf{NO\_EXAM} as ${\bm m}=(m_1,\dots,m_{197})^\top$) in the $197$ villages individually. The Loaloa dataset also contains four explanatory variables, including the elevation (${\bm x}_1$ in $km$), the mean Normalized Difference Vegetation Index (NDVI) (${\bm x}_2$), the max NDVI (${\bm x}_3$), and the min NDVI (${\bm x}_4$) values at the village level. 

\begin{table}
\caption{\label{tab:loa loa data spatial analysis} Estimates of ${\bm\beta}$ and ${\bm\omega}$ in the logistic SGLMMs for the {\it Loaloa} data given by our proposed and the previous PQL and LA methods, where the standard error of the estimates are given in the parentheses.}
\begin{center}
\begin{tabular}{cccc}\hline
         & \multicolumn{3}{c}{Methods} \\\cline{2-4}
 Estimates  & Proposed  & PQL & LA \\\hline
     & \multicolumn{3}{c}{${\mathscr M}_0:\breve{\bm\eta}=\beta_0+{\bm x}_1\beta_1+{\bm x}_2\beta_2+{\bm x}_3\beta_3$}  \\
 $\beta_0$ & $-9.39(1.42)$ & $-9.23(1.45)$ & $-9.58(1.45)$\\
 $\beta_1$& $-0.91(0.32)$& $-0.91(0.33)$ & $-0.94(0.33)$\\
 $\beta_2$ & \hspace{7pt}$7.45(2.20)$ & \hspace{7pt}$7.40(2.23)$ &\hspace{5pt} $7.73(2.24)$  \\
 $\beta_3$ & \hspace{7pt}$5.07(1.86)$ & \hspace{7pt}$4.89(1.87)$ &\hspace{5pt}  $5.10(1.89)$\\
 $\omega_1$ &  \hspace{7pt}$0.58(0.06)$ & \hspace{7pt}$0.61({\rm NA})$ &\hspace{5pt} $0.59({\rm NA})$\\
 $\omega_2$ &  \hspace{7pt}$2.17(0.63)$ & \hspace{7pt}$1.88({\rm NA})$ &\hspace{5pt} $2.10({\rm NA})$\\       
     & \multicolumn{3}{c}{${\mathscr M}_1: \breve{\bm\eta}=\beta_0+{\bm x}_1\beta_1+{\bm x}_2\beta_2$}  \\
 $\beta_0$ & $-6.58(1.04)$ &  $-6.54(1.07)$ & $-6.76(1.06)$  \\
 $\beta_1$ & $-1.09(0.32)$ &  $-1.08(0.32)$ & $-1.12(0.33)$ \\
  $\beta_2$ & \hspace{3pt}$10.19(1.98)$ & \hspace{3pt}$10.06(2.01)$ & \hspace{3pt}$10.50(2.02)$ \\
   $\omega_1$ & \hspace{7pt}$0.62(0.06)$ & \hspace{7pt}$0.65({\rm NA})$ & \hspace{7pt}$0.63({\rm NA})$\\
    $\omega_2$ & \hspace{7pt}$1.87(0.57)$ & \hspace{7pt}$1.61({\rm NA})$ & \hspace{7pt}$1.81({\rm NA})$ \\
         & \multicolumn{3}{c}{${\mathscr M}_2:\breve{\bm\eta}=\beta_0+{\bm x}_1\beta_1+{\bm x}_3\beta_3$}  \\
 $\beta_0$ & $-8.11(1.42)$ & $-7.96(1.44)$ & $-8.27(1.44)$ \\
 $\beta_1$ & $-0.67(0.32)$ & $-0.69(0.32)$ & $-0.70(0.33)$  \\
  $\beta_3$ &  \hspace{7pt} $7.90(1.66)$ & \hspace{7pt}$7.70(1.67)$ & \hspace{10pt}$8.04(1.69)$ \\
   $\omega_1$ & \hspace{7pt} $0.61(0.06)$ & \hspace{7pt}$0.64({\rm NA})$ &\hspace{5pt} $0.63({\rm NA})$ \\
    $\omega_2$ & \hspace{7pt} $1.94(0.58)$ & \hspace{7pt}$1.68({\rm NA})$ &\hspace{5pt} $1.88({\rm NA})$ \\  
             & \multicolumn{3}{c}{${\mathscr M}_3:\breve{\bm\eta}=\beta_0+{\bm x}_2\beta_2+{\bm x}_3\beta_3$}  \\
 $\beta_0$ & $-10.17(1.42)$ & $-10.05(1.45)$ & $-10.41(1.45)$ \\
 $\beta_2$ & \hspace{11pt}$6.10(2.18)$ & \hspace{11pt} $6.15(2.22)$ &\hspace{11pt}$6.37(2.23)$\\
  $\beta_3$ & \hspace{11pt}$6.16(1.84)$ &\hspace{11pt} $5.97(1.86)$ &\hspace{11pt} $6.23(1.87)$ \\
   $\omega_1$ &\hspace{9pt} $0.60(0.06)$ &\hspace{9pt} $0.63({\rm NA})$ & \hspace{9pt} $0.62({\rm NA})$ \\
    $\omega_2$ &\hspace{9pt} $2.07(0.60)$ &\hspace{9pt} $1.80({\rm NA})$ &\hspace{9pt} $2.00({\rm NA})$  \\        
                 & \multicolumn{3}{c}{${\mathscr M}_4: \breve{\bm\eta}=\beta_0+{\bm x}_1\beta_1+{\bm x}_2\beta_2+{\bm x}_3\beta_3+{\bm x}_4\beta_4$}  \\
 $\beta_0$ & $-9.40(1.42)$  &$-9.24(1.45)$ &  $-9.60(1.45)$\\
 $\beta_1$ & $-0.91(0.32)$  &$-0.90(0.33)$ &  $-0.93(0.33)$ \\
 $\beta_2$ &\hspace{7pt} $7.61(2.27)$  &\hspace{7pt}$7.52(2.31)$ &\hspace{7pt} $7.86(2.31)$ \\
  $\beta_3$ & \hspace{9pt}$5.05(1.86)$ &\hspace{9pt}$4.87(1.88)$ & \hspace{11pt}$5.08(1.89)$\\
  $\beta_4$ & $-0.32(1.19)$ &$-0.23(1.19)$ &  $-0.28(1.21)$ \\
   $\omega_1$ &\hspace{5pt} $0.58(0.06)$ & \hspace{5pt}$0.60({\rm NA})$ &\hspace{5pt} $0.69({\rm NA})$\\
    $\omega_2$ & \hspace{9pt}$2.20(0.63)$ & \hspace{5pt}$1.90({\rm NA})$ & \hspace{7pt}$2.13({\rm NA})$ \\\hline         
\end{tabular}
\end{center}
\end{table}

Before applying our method, we used $\log[{\bm y}/({\bm m}-{\bm y})]$ adopted by the \textsf{PrevMap} package to look at the pattern of the disease briefly. We carried out a Gaussian kriging method by treating $\log[{\bm y}/({\bm m}-{\bm y})]$ as the variable of interest. We calculated the distance matrix between the $197$ villages. We interpolated 
the logistic values at the unobserved villages. We changed the logistic values back to the probabilities for the infections of {\it Loa loa} in the entire study region (Figure~\ref{fig:loaloa locations}). It partially reflected a global pattern of {\it Loa Loa} infections in the study region. 

We applied the proposed and the previous PQL and LA methods to the {\it Loaloa} data. We devised a few logistic SGLMMs by assuming ${\bm y}|{\bm\gamma}\sim Bin({\bm m},{\bm\pi})$ with
\begin{equation}
\label{eq:formulations models in loa loa}
\log{{\bm\pi}\over {\bm 1}-{\bm\pi}}=\breve{\bm\eta}+{\bm\gamma}.
\end{equation}
The prior distribution for ${\bm\gamma}$ is specified by~\eqref{eq:distribution of random effects} with the variance component matrix ${\bf D}_{\bm\omega}$ specified by the Mat\'ern family with $\omega_3=0.5$ given by~\eqref{eq:matern family}. We found that the choice was appropriate by investigating the influence of $\omega_3$.

We considered a few options of the fixed effects components $\breve{\bm\eta}$ in~\eqref{eq:formulations models in loa loa}. We displayed the options of $\breve{\bm\eta}$ in Table~\ref{tab:loa loa data spatial analysis}. After comparing the significance of explanatory variables, we found that ${\bm x}_1$ (i.e., elevation), ${\bm x}_2$ (i.e., mean NDVI), and ${\bm x}_3$ (i.e.,  max NVDI) were significant but ${\bm x}_4$ (i.e., min NDVI) was not, leading to ${\mathscr M}_0$ as the final model. The standard errors of $\hat{\bm\beta}$ and $\hat{\bm\omega}$ in our method were derived based on the sample Fisher Information given by~\eqref{eq:fisher information}. The estimates in the PQL and LA methods were derived by the \textsf{spaMM} packages of $\textsf{R}$. The \textsf{spaMM} did not provide the standard errors of $\hat{\bm\omega}$. They were not available in the output of the corresponding models.

To derive ${\mathscr M}_0$, we investigated three candidate models by removing one of the explanatory variables, leading to ${\mathscr M}_1$, ${\mathscr M}_2$, and ${\mathscr M}_3$ in Table~\ref{tab:loa loa data spatial analysis}. We used the likelihood ratio statistic $\Lambda_\psi$ given by~\eqref{eq:likelihood ratio statistics}, the score statistic $S_{\psi}$ given by~\eqref{eq:score statistic}, and the generalized Wald statistic $GW_{\psi}$ given by~\eqref{eq:generalized wald statistic} to measure the differences between ${\mathscr M}_0$ and the other candidate models, individually. We had $\Lambda_\psi=6.76,11.49,6.61$ with $p$-values $0.0093,0.0007,0.0106$, $S_\psi=6.36,10.50,7.58$ with $p$-values $0.012,0.0012,0.0060$, and $GW_\psi=6.78,9.98,7.13$ with $p$-values $0.009,0.0015,0.0076$, respectively. In addition, we evaluated another candidate model by adding ${\bm x}_4$ to ${\mathscr M}_0$, leading to ${\mathscr M}_4$ in Table~\ref{tab:loa loa data spatial analysis}. We obtained $\Lambda_\psi=0.194$ with $p$-value $0.660$, $S_{\psi}=0.069$ with $p$-value $0.793$, and $GW_{\psi}=0.076$ with $p$-value $0.783$, respectively. The likelihood ratio statistic values in the PQL method for the difference between ${\mathscr M}_0$ and ${\mathscr M}_1,{\mathscr M}_2,{\mathscr M}_3$ were $9.51,14.72,7.45$ with $p$-values $0.002,0.0001,0.0063$, respectively. The likelihood ratio statistics in the LA method were $6.84,11.61,8.06$ with $p$-values $0.009,0.0007,0.0045$, respectively. The likelihood ratio statistic in the PQL method for the difference between ${\mathscr M}_0$ and ${\mathscr M}_4$ was $0.050$ with $p$-value $0.823$. The likelihood ratio statistic in the LA method was $0.051$ with $p$-value $0.821$. The score and generalized Wald statistics were unavailable in PQL and the LA methods. We concluded ${\mathscr M}_0$ the best model in our proposed and the previous PQL and LA methods.

We found that the estimates of ${\bm\beta}$ and ${\bm\omega}$ in our proposed method were close to those in the previous PQL and LA methods, implying that the approximations of the likelihood function given by the PQL and LA methods were precise in the application. We examined the reason and found that it was caused by the regression effects contained in the fixed effect components. The standard errors of the explanatory variables in the {\it Loaloa} data were ${\rm SE}({\bm x}_1)=0.397$, ${\rm SE}({\bm x}_2)=0.061$, ${\rm SE}({\bm x}_3)=0.057$, and ${\rm SE}({\bm x}_4)=0.063$. The magnitudes were lower than those adopted in our simulation studies. In this case, the PQL and the LA methods can be treated as precise in estimating ${\bm\beta}$ and ${\bm\omega}$. Our method provides a base to measure the accuracy and precision of previous methods for estimating the parameters and hyperparameters.

\section{Conclusion}
\label{sec:conclusion}

The exact MLE for GLMMs is a long-standing problem in the literature. We solve the exact MLE problem but not the exact likelihood problem. The approach is that computing the exact MLE only needs the gradient vector of the likelihood function but not the likelihood function itself. Based on the properties of the gradient vector of the likelihood function, we construct a sequence of objective functions to result the exact MLE. The previous LA or Monte Carlo-based methods cannot achieve this. We do not treat any intractable integrals numerically, implying that intractable integrals are not an issue in our method. We expect that our idea can be implemented into broader Bayesian frameworks when intractable integrals are involved. This is left to future research. 

\appendix

\section{Proofs}
\label{sec:proofs}

\noindent
{\bf Proof of Lemma~\ref{lem:normal density posterior identity}.} The first statement $\hat{\bm\gamma}_{{\bm\beta}{\bm\omega}}={\bm v}_{{\bm\beta}{\bm\omega},{\bm 0}{\bm 0}}$ is implied by~\cite{zhang2024}. For the second statement, we simplify our notations as $\tilde{\bm y}=\tilde{\bm y}_{{\bm\beta}\hat{\bm\gamma}_{{\bm\beta}{\bm\omega}}}$,  ${\bf W}={\bf W}_{{\bm\beta}\hat{\bm\eta}_{{\bm\beta}{\bm\omega}}}$, ${\bf D}={\bf D}_{{\bm\omega}+{\bm\delta}}$, ${\bm v}={\bm v}_{{\bm\beta}{\bm\omega},{\bm\alpha}{\bm\delta}}$, ${\bf V}={\bf V}_{{\bm\beta}{\bm\omega},{\bm\delta}}$, and ${\bf R}={\bf R}_{{\bm\beta}{\bm\omega},{\bm\delta}}$. We express logarithm of the left-hand of~\eqref{eq:normal density identity for initial modified objective function} as
$$
\log[\varphi(\tilde{\bm y};{\bf X}{\bm\beta}+{\bf X}{\bm\alpha}+{\bf Z}{\bm\gamma},{\bf W}^{-1})\varphi({\bm\gamma};{\bf 0},{\bf D})]=-{n+r\over 2}\log(2\pi)+T_1+T_2+T_3+T_4,
$$
where $T_1=-(1/2)[\log|\det({\bf W}^{-1})|+\log|\det({\bf D})|]$, $T_2=-(1/2)(\tilde{\bm y}-{\bf X}{\bm\beta}-{\bf X}{\bm\alpha}){\bf W}(\tilde{\bm y}-{\bf X}{\bm\beta}-{\bf X}{\bm\alpha})$, $T_3={\bm\gamma}^\top{\bf Z}^\top {\bf W}(\tilde{\bm y}-{\bf X}{\bm\beta}-{\bf X}{\bm\alpha})$, and $T_4=-(1/2){\bm\gamma}^\top ({\bf D}^{-1}+{\bf Z}^\top{\bf W}{\bf Z}^\top){\bm\gamma}$.  We express the right-hand side of~\eqref{eq:normal density identity for initial modified objective function} as
$$
\log[\varphi({\bm\gamma}; {\bm v},{\bf V})\varphi(\tilde{\bm y}; {\bf X}{\bm\beta}+{\bf X}{\bm\alpha},{\bf R})]=-{n+r\over 2}\log(2\pi)+\tilde T_1+\tilde T_2+\tilde T_3+\tilde T_4,
$$
where $\tilde T_1= -(1/2)[\log|\det({\bf R})|+\log|\det({\bf D}-{\bf D}{\bf Z}^\top{\bf R}^{-1}{\bf Z}{\bf D})|]$,  $\tilde T_2= -(1/2)(\tilde{\bm y}-{\bf X}{\bm\beta}-{\bf X}{\bm\alpha})^\top\{{\bf R}^{-1}-{\bf R}^{-1}{\bf Z}{\bf D}({\bf D}-{\bf D}{\bf Z}^\top{\bf R}^{-1}{\bf Z}{\bf D})^{-1}{\bf D}{\bf Z}^\top{\bf R}^{-1}\}(\tilde{\bm y}-{\bf X}{\bm\beta}-{\bf X}{\bm\alpha})$, $\tilde T_3={\bm\gamma}^\top\{({\bf D}-{\bf D}{\bf Z}^\top{\bf R}^{-1}{\bf Z}{\bf D})^{-1}{\bf D}{\bf Z}^\top{\bf R}^{-1}\}(\tilde{\bm y}-{\bf X}{\bm\beta}-{\bf X}{\bm\alpha})$, and  $\tilde T_4= -(1/2){\bm\gamma}^\top({\bf D}-{\bf D}{\bf Z}^\top{\bf R}^{-1}{\bf Z}{\bf D})^{-1}{\bm\gamma}$. We want to show $T_1=\tilde T_1$, $T_2=\tilde T_2$, $T_3=\tilde T_3$, $T_4=\tilde T_4$. Using the Woodbury matrix identity, we have $[
{\bf D}^{-1}+{\bf Z}^\top{\bf W}{\bf Z}]^{-1}={\bf D}-{\bf D}{\bf Z}^\top({\bf W}^{-1}+{\bf Z}{\bf D}{\bf Z}^\top)^{-1}{\bf Z}{\bf D}$, leading to ${\bf D}^{-1}+{\bf Z}^\top{\bf W}{\bf Z}=[{\bf D}-{\bf D}{\bf Z}^\top({\bf W}^{-1}+{\bf Z}{\bf D}{\bf Z}^\top)^{-1}{\bf Z}{\bf D}]^{-1}$. We obtain $T_4=\tilde T_4$. Still using the Woodbury matrix identity, we have
$$\eqalign{
&({\bf W}^{-1}+{\bf Z}{\bf D}{\bf Z}^\top)^{-1}+({\bf W}^{-1}+{\bf Z}{\bf D}{\bf Z}^\top)^{-1}{\bf Z}{\bf D}\cr
&\hspace{2cm}[{\bf D}-{\bf D}{\bf Z}^\top({\bf W}^{-1}+{\bf Z}{\bf D}{\bf Z}^\top)^{-1}{\bf Z}{\bf D}]^{-1}{\bf D}{\bf Z}^\top({\bf W}^{-1}+{\bf Z}{\bf D}{\bf Z}^\top)^{-1}\cr
=&({\bf W}^{-1}+{\bf Z}{\bf D}{\bf Z}^\top)^{-1}+({\bf W}^{-1}+{\bf Z}{\bf D}{\bf Z}^\top)^{-1}{\bf Z}{\bf D}{\bf Z}^\top{\bf W}\cr
=&{\bf W},
}$$
leading to $T_2=\tilde T_2$. Also by the Woodbury matrix identity, we have
$$\eqalign{
&[{\bf D}-{\bf D}{\bf Z}^\top({\bf W}^{-1}+{\bf Z}{\bf D}{\bf Z}^\top)^{-1}{\bf Z}{\bf D}]^{-1}{\bf D}{\bf Z}^\top({\bf W}^{-1}+{\bf Z}{\bf D}{\bf Z}^\top)^{-1}\cr
=&({\bf D}^{-1}+{\bf Z}^\top{\bf W}{\bf Z}){\bf D}{\bf Z}^\top({\bf W}^{-1}+{\bf Z}{\bf D}{\bf Z}^\top)^{-1}\cr
=& {\bf Z}^\top ({\bf W}^{-1}+{\bf Z}{\bf D}{\bf Z}^\top)^{-1}+{\bf Z}^\top{\bf W}{\bf Z}{\bf D}{\bf Z}^\top({\bf W}^{-1}+{\bf Z}{\bf D}{\bf Z}^\top)^{-1}\cr
=&{\bf Z}^\top {\bf W}{\bf W}^{-1}({\bf W}^{-1}+{\bf Z}{\bf D}{\bf Z}^\top)^{-1}+{\bf Z}^\top{\bf W}{\bf Z}{\bf D}{\bf Z}^\top({\bf W}^{-1}+{\bf Z}{\bf D}{\bf Z}^\top)^{-1}\cr
=&{\bf Z}^\top{\bf W},
}$$
leading to $T_3=\tilde T_3$. For $T_1$ and $\tilde T_1$, we study the determinant of 
$${\bf C}=\left( \begin{array}{cc}{\bf D} & {\bf D} {\bf Z}^\top \cr {\bf Z}{\bf D}  & {\bf W}^{-1}+{\bf Z}{\bf D}{\bf Z}^\top   \end{array}  \right).$$
We have
$$\eqalign{
\det({\bf C})=&\det \left[ \left(  \begin{array}{cc}{\bf D} & {\bf D} {\bf Z}^\top \cr {\bf Z}{\bf D}  & {\bf W}^{-1}+{\bf Z}{\bf D}{\bf Z}^\top   \end{array}  \right)\left(\begin{array}{cc} {\bf I} & -{\bf Z}^\top \cr {\bf 0} & {\bf I} \end{array}   \right) \right]\cr
=&\det\left(\begin{array}{cc}{\bf D} & {\bf 0} \cr {\bf Z}{\bf D}_{\bm\omega} & {\bf W}^{-1} \end{array}   \right) \cr
=&\det({\bf D})\det({\bf W}^{-1})
}$$
and 
$$\eqalign{
\det({\bf C})=&\det \left[\left(\begin{array}{cc} {\bf I} &  -{\bf D}{\bf Z}^\top({\bf W}^{-1}+{\bf Z}{\bf D}{\bf Z}^\top)^{-1} \cr {\bf 0} & {\bf I} \end{array}   \right)  \left(  \begin{array}{cc}{\bf D} & {\bf D}{\bf Z}^\top \cr {\bf Z}{\bf D}  & {\bf W}^{-1}+{\bf Z}{\bf D}{\bf Z}^\top   \end{array}  \right)\right]\cr
=&\det\left( \begin{array}{cc} {\bf D}-{\bf D}{\bf Z}^\top({\bf W}^{-1}+{\bf Z}{\bf D}{\bf Z}^\top)^{-1}{\bf Z}{\bf D}& {\bf 0} \cr {\bf Z}{\bf D} & {\bf W}^{-1}+{\bf Z}{\bf D}{\bf Z}^\top \cr  \end{array}   \right)\cr
=&\det[ {\bf D}-{\bf D}{\bf Z}^\top({\bf W}^{-1}+{\bf Z}{\bf D}{\bf Z}^\top)^{-1}{\bf Z}{\bf D}]\det({\bf W}^{-1}+{\bf Z}{\bf D}{\bf Z}^\top),
}$$
leading to $T_1=\tilde T_1$. We conclude.  \qed

\noindent
{\bf Proof of Theorem~\ref{thm:gradient property of modified working objective function for MLE}.} Based on ${\rm E}_{{\bm\beta}{\bm\omega}}({\bm\gamma}|{\bm y})= {\bm v}_{{\bm\beta}{\bm\omega},{\bm 0}{\bm 0}}={\bf D}_{\bm\omega}{\bf Z}^\top{\bf R}_{{\bm\beta}{\bm\omega},{\bm 0}}^{-1} (\tilde{\bm y}_{{\bm\beta}\hat{\bm\gamma}_{{\bm\beta}{\bm\omega}}}-{\bf X}{\bm\beta})$ and ${\rm cov}_{{\bm\beta}{\bm\omega}}({\bm\gamma}|{\bm y})= {\bf V}_{{\bm\beta}{\bm\omega},{\bm 0}}={\bf D}_{\bm\omega}-{\bf D}_{\bm\omega}{\bf Z}^{-1}{\bf R}_{{\bm\beta}{\bm\omega},{\bm 0}}^{-1}{\bf Z}{\bf D}_{\bm\omega}$ given by~\cite{zhang2024} with the properties of the expectation of the normal likelihood given by~\cite{zhang2019}, we have 
$${\rm E}_{{\bm\beta}{\bm\omega}}[\varphi_{{\bm\beta}{\bm\omega}}({\bm\gamma}; {\bm v}_{{\bm\beta}{\bm\omega},{\bm\alpha}{\bm\delta}},{\bf V}_{{\bm\beta}{\bm\omega},{\bm\delta}})|{\bm y}]\le {\rm E}_{{\bm\beta}{\bm\omega}}[\varphi({\bm\gamma}; {\bm v}_{{\bm\beta}{\bm\omega},{\bm 0}{\bm 0}},{\bf V}_{{\bm\beta}{\bm\omega},{\bm 0}})|{\bm y}]$$
and the inequality holds iff ${\bm\alpha}={\bm 0}$ and ${\bm\delta}={\bm 0}$. Applying the properties of ${\mathcal A}$ given by~\eqref{eq:solution set of MLE for score equation likelihood} and~\eqref{eq:identity for gradients of beta of the loglikelihood and initial working objective function of MLE}, we conclude. \qed

\noindent
{\bf Proof of Corollary~\ref{cor:solution of local}.} For the first statement, the final ${\bm\alpha}^{(t)}$ and ${\bm\delta}^{(t)}$ satisfies ${\bm\alpha}^{(t)}={\bf 0}$ and ${\bm\delta}^{(t)}={\bm 0}$, leading to the final $\dot\psi^{(t)}({\bm 0},{\bm 0})={\bm 0}$, implying that the final $({\bm\beta}^{(t)\top},{\bm\omega}^{(t)\top})^\top\in{\mathcal A}$. For the second statement, if ${\bm\beta}^{(t-1)}=\check{\bm\beta}$ and ${\bm\beta}^{(t-1)}=\check{\bm\omega}$ for some $(\check{\bm\beta}^\top,\check{\bm\omega}^\top)^\top\in{\mathcal A}$ in the previous iteration, then there is $\dot\psi^{(t)}({\bm 0},{\bm 0})={\bm 0}$, leading to ${\bm\beta}^{(t)}={\bm\beta}^{(t-1)}=\check{\bm\beta}$ and ${\bm\omega}^{(t)}={\bm\omega}^{(t-1)}=\check{\bm\omega}$ in the current iteration. We conclude. \qed

\end{document}